\documentclass[traditabstract]{aa}
\pdfoutput=1
\usepackage{txfonts}
\usepackage{longtable}
\usepackage{subfig}
\usepackage{url}
\usepackage{graphicx}
\usepackage{natbib}
\bibpunct{(}{)}{;}{a}{}{,}

\begin{document}

  \title{T-RaMiSu: The Two-meter Radio Mini Survey}
  \subtitle{I. The Bo\"{o}tes Field}
  \titlerunning{T-RaMiSu: The Two-meter Radio Mini Survey. I.}
  
  \author{  W.~L.~Williams\inst{\ref{inst1},\ref{inst3}}
     \and   H.~T.~Intema\inst{\ref{inst2}}\fnmsep\thanks{Jansky Fellow of the National Radio Astronomy Observatory} 
     \and   H.~J.~A.~R\"{o}ttgering\inst{\ref{inst1}} }
  
  \institute{ Leiden Observatory, Leiden University, P.O.Box 9513, NL-2300 RA, Leiden, The Netherlands \label{inst1}\\
              \email{wwilliams@strw.leidenuniv.nl}
        \and  Netherlands Institute for Radio AStronomy (ASTRON), PO Box 2, 7990AA Dwingeloo, The Netherlands \label{inst3}
        \and  National Radio Astronomy Observatory, 520 Edgemont Road, Charlottesville, VA 22903-2475, USA\label{inst2} }
        
  \date{Received  /  Accepted }

  \abstract{
  We present wide area, deep, high-resolution $153$~MHz GMRT observations of the NOAO Bo\"{o}tes field, adding to the extensive, multi-wavelength data of this region. The observations, data reduction, and catalogue construction and description are described here. The {seven} pointings produced a final mosaic covering $30$~square degrees with a resolution of $25\arcsec$. The $rms$ noise is $2$~mJy~beam$^{-1}$ in the centre of the image, rising to $4-5$~mJy~beam$^{-1}$ on the edges, with an average of $3$~mJy~beam$^{-1}$. {Seventy-five} per cent of the area has an $rms < 4$~mJy~beam$^{-1}$. The extracted source catalogue contains $1289$\ sources detected at $5\sigma$, of which $453$ are resolved. We estimate the catalogue to be $92$~per~cent reliable and $95$~per~cent complete at an integrated flux density limit of $14$~mJy.  The flux densities and astrometry have been corrected for systematic errors. We calculate the differential source counts {which are in good agreement with those in the literature and provide an important step forward in quantifying the source counts at these low frequencies and low flux densities}. The GMRT $153$~MHz sources have been matched to the $1.4$~GHz NVSS  and $327$~MHz WENSS catalogues and spectral indices were derived.  }


\maketitle

\titlerunning{Low Frequency Observations of the Bo\"{o}tes Field}

\section{Introduction}
Deep low-frequency radio surveys provide unique data which will help resolve many questions related to the formation and evolution of massive galaxies, quasars and galaxy clusters. Until now, such surveys have largely been limited by the corrupting influence of the ionosphere on the visibility data, but new techniques allow for the correction for these effects \citep[e.g.][]{Cotton+2004,Intema+2009}. Recently deep ($0.7-2$~mJy~beam$^{-1}$) images have been made, in particular with the Giant Metrewave Radio Telescope \citep[GMRT, e.g.][]{Ananthakrishnan2005} at $153$~MHz \citep[e.g.][]{IshwaraChandraMarathe2007,Sirothia+2009,IshwaraChandra+2010}. These observations can be used to study:

 \textit{Luminous radio sources at $z>4$ -- }  High redshift radio galaxies \citep[HzRGs, e.g.][]{MileyDeBreuck2008} provide a unique way to study the evolution of the  most massive galaxies in the Universe. One of the most efficient ways of identifying these sources is to search for ultra-steep spectrum (USS) radio sources with $\alpha \lesssim -1$, $S_{\nu} \propto \nu^{\alpha}$ \citep{Rottgering+1997,deBreuck+2002}. Low frequency observations provide an easy way of identifying USS sources and extending these observations to lower flux density limits increases the distance to which these HzRGs can be identified. Surveying larger areas increases the probability of locating these rare sources.


 \textit{Distant starburst galaxies -- }
 The local radio-IR correlation for star forming galaxies is very tight, and seems to hold at high redshift
\citep{Kovacs+2006}. However, the physical processes involved are poorly understood and only models that carefully fine-tune the time scales for the heating of the dust, the formation of supernovae, and the acceleration, diffusion and decay of the relativistic electrons can reproduce the correlation. The low-frequency spectral shape of galaxies reveals information about the amount of free-free absorption and relating this to the dust content, size, mass, total amount of star formation and environment of the galaxies will further constrain the radio-IR models. {To date, however, few galaxies have been well studied at low frequencies and those that have, show a diverse range of spectral shapes \citep[e.g.][]{Clemens+2010}. }

 \textit{Faint peaked spectrum sources -- }
 Young radio-loud AGN are ideal objects to study the onset and early evolution of classical double radio
sources. They usually have synchrotron self-absorbed spectra and compact radio morphologies. Relative number statistics have indicated that these radio sources must be significantly more powerful at young ages, which may be preceded by a period of luminosity increase \citep[e.g.][]{Snellen+2003}. Multi-epoch VLBI observations of individual Gigahertz Peaked Spectrum and Compact Symmetric Objects indicate dynamical ages in the range of a few hundred to a few thousand years \citep[e.g][]{PolatidisConway2003}. {Since the peak of these sources shifts to lower observed frequencies at higher redshift, low frequency observations, combined with multi-wavelength data,} can identify these faint peakers and establish whether they are less luminous or at very high redshift and have different host properties (masses, starformation rates). 

 \textit{The accretion modes of radio sources -- }
Radio galaxies and radio loud quasars have been studied extensively in order to reveal the details of the relationship between Active Galactic Nuclei (AGN) and their host galaxies, in particular how their interaction affects their evolution. The expanding jets of radio-loud AGN provide a mechanism for the transfer of energy to the intracluster medium and prevent the catastrophic cooling and formation of too-massive elliptical galaxies  \citep{Fabian+2006,Best+2006MNRAS,Best+2007MNRAS,Croton+2006,Bower+2006}, but the accretion and feedback processes and how they evolve over cosmic time are not fully understood.  It is known that the fraction of massive galaxies which are radio-loud  at $z \sim 0.5$ is about the same as{ observed locally \citep[$z \sim 0.1$,][]{Best+2005}}, while for less massive galaxies ($< 10^{10.5} M_{\odot} $), it is an order of magnitude larger. Studies of these AGN show two different types: a ``hot'' mode where radiatively inefficient accretion occurs from hot halo gas onto massive galaxies, and  a ``cold'' mode where cold gas from major mergers drives high accretion rates. The strong evolution in the radio luminosity function is thus a result of less massive galaxies experiencing more mergers and being more active at high $z$. A full understanding of the different AGN populations, their distribution in luminosity and host galaxy properties, and particularly their cosmic evolution, is important for AGN and galaxy evolutionary models. Differences in their host galaxy populations will provide insight into the triggering mechanisms for radio activity as well as the effect of radio feedback.



In this paper we present wide, deep, high-resolution observations of the NOAO Bo\"{o}tes extra-galactic field at $153$~MHz taken with the GMRT. An initial, very deep, $\sim\!1$~mJy~beam$^{-1}$ $rms$, $153$~MHz GMRT map of this field was presented by \cite{Intema+2011}. Here we present additional pointings around this map effectively tripling the size of the surveyed area at a slightly higher noise level. The Bo\"{o}tes field is part of the NOAO Deep Wide Field Survey \citep[NDWFS;][]{JannuziDey1999} and covers $\sim\!9$~deg$^2$ in the optical and near infra-red $B_W$, $R$, $I$ and $K$ bands. There is a wealth of additional complementary data available for this field, including 
X-ray \citep{Murray+2005,Kenter+2005}, UV \citep[GALEX;][]{Martin+2003}, and mid infrared \citep{Eisenhardt+2004,Martin+2003}. The region has also been surveyed at radio wavelengths with the WSRT at 1.4 GHz \citep{deVries+2002}{}, the VLA at 1.4 GHz \citep{Higdon+2005} and 325 MHz \citep{Croft+2008}. Recently, the AGN and Galaxy Evolution Survey (AGES) has provided redshifts for $23\,745$ galaxies and AGN across $7.7$~deg$^2$ of the Bo\"{o}tes field \citep{Kochanek+2012}.
This unique rich multiwavelength dataset, combined with the new low frequency radio data presented here, will be valuable in improving our understanding of the above-mentioned key topics in astrophysics.

The observations presented here are the first part of the Two-meter Radio Mini Survey (T-RaMiSu), consisting of two $153$~MHz mosaics of similar area and depth. The second mosaic, centered on the galaxy cluster Abell 2256, will be presented by Intema et. al \citetext{in prep}.

This paper is structured as follows. In Sect.~\ref{sect:obsred} we describe the GMRT observations of the extended region around the NOAO Bo\"{o}tes field. We describe the techniques employed to achieve the deepest possible images. Our data reduction relies on the ionospheric calibration with the SPAM package \citep{Intema+2009}. In Sect.~\ref{sect:sources} we describe the source detection method and the compilation of a source catalogue. This section also includes a discussion of the completeness and reliability of the catalogue and an analysis of the quality of the catalogue. The source counts and spectral index distributions are presented in Sect. \ref{sect:analysis}. Finally, Sect. \ref{sect:concl} summarises and concludes {this work}.

\section{Observations and Data Reduction}
\label{sect:obsred}

\subsection{Observations}
The central Bo\"{o}tes field was previously observed with the GMRT from $3-4$ June 2005 \citep{Intema+2011}. We use the data from a single day of this observing run, combined with new observations of six flanking fields taken during $3-6$ June 2006 with the GMRT at $153$~MHz. {Data from the first day only of the first observing run, 3 June, was used as the RFI situation was marginally better on this day and the length of a single day's observation, $359$~min, compares well with that of the new observations of the flanking fields, $205$~min, which leads to a more uniform mosaic.} Table \ref{tab:observations} lists the observational parameters used, highlighting any differences between the two sets of observations. The flanking fields are arranged on a hexagonal grid with a radius of $110\arcmin$ just beyond the half power point of the primary beam of the GMRT at $153$~MHz ($\theta_{\mathrm{FWHM}} \sim\!3^{\circ}$); Table \ref{tab:pointings} gives the central coordinates of each pointing. Typically $26-27$ of the $30$ antennas were available during each observing run. 3C\,48 and 3C\,286 were observed as phase, bandpass and flux density calibrators. For each of the four days, the target fields were observed in sets of $\sim\!4.5$~min each, followed by a calibrator observation (3C\,286) of $\sim\!4.5$~min. 3C\,48  was observed at the beginning or end of each day for $\sim\!20-30$\ min. The frequent ($\sim\!30$~min interval) calibrator observations of 3C\,286 provide a means to track changes in the GMRT system, RFI and ionospheric conditions, and flux density scale. The short target field observations spread over each night of observing provides fairly uniform $uv$-coverage.

\begin{table}[tp]
 \begin{center}
 \caption{GMRT observation parameters for the Bo\"{o}tes field.}
 \label{tab:observations}
\begin{tabular}{lll}
\hline 
Parameter &  \multicolumn{1}{c}{Central} &  \multicolumn{1}{c}{Flanking} \\
\hline
Observation Dates & 3 June 2005 & 3-6 June 2006 \\
Pointings & Bo\"{o}tes & Bo\"{o}tes A-F \\
Primary Calibrator & 3C\,48 & 3C\,48 \\
~~Total Time on Calibrator & $20$~min & $51$~min \\
Secondary Calibrator & 3C\,286 & 3C\,286 \\
~~Total Time on Calibrator & $9\times10$~min  & $10\times4.5$~min (per day) \\
~~Cadence & $50$~min & $30$~min \\
Total Time on Target & $359$~min & $205$~min (per pointing) \\
\hline
Integration time & \multicolumn{2}{c}{$16.9$~s} \\
Polarisations & \multicolumn{2}{c}{RR,LL} \\
Channels & \multicolumn{2}{c}{$128$ } \\
Channel Width & \multicolumn{2}{c}{$62.5$~kHz} \\
Total Bandwidth & \multicolumn{2}{c}{$8.0$~MHz} \\
Central Frequency & \multicolumn{2}{c}{$153$~MHz}\\
\hline
 \end{tabular}
 \end{center}
\end{table}

\begin{table}
 \begin{center}
 \caption{Pointing centres of the Bo\"{o}tes central and flanking fields.}
 \label{tab:pointings}
\begin{tabular}{lcc}
\hline
Field & RA & DEC \\ 
 & (J2000) & (J2000) \\ 
\hline
Bo\"{o}tes & 14:32:05.75 & +34:16:47.5 \\
Bo\"{o}tes A & 14:32:05.75 & +36:06:47.5 \\
Bo\"{o}tes B & 14:24:19.53 & +35:10:52.5 \\
Bo\"{o}tes C & 14:24:29.58 & +33:20:54.5 \\
Bo\"{o}tes D & 14:32:05.75 & +32:26:47.5 \\
Bo\"{o}tes E & 14:39:41.92 & +33:20:54.5 \\
Bo\"{o}tes F & 14:39:51.97 & +35:10:52.5 \\
\hline
 \end{tabular}
 \end{center}
\end{table}

\subsection{Data Reduction}
The data for the central pointing was re-reduced in the same manner as the new flanking fields in order to allow for consistent integration into a single mosaic.
The data reduction consisted of two stages: ``traditional calibration'' followed by directional-dependent ionospheric phase calibration, both of which were implemented in Python using the ParselTongue \citep{Kettenis+2006} interface to the Astronomical Image Processing System package \citep[AIPS;][]{Greisen1998}. Ionospheric calibration was done with the ``Source  Peeling and Atmospheric Modelling'' ParselTongue-based Python module \citep[SPAM;][]{Intema+2009}.

The data for each day were calibrated separately. The flux density scale was set and initial amplitude, phase and bandpass calibration were done using 3C\,48. 3C\,48 is brighter than 3C\,286 and provides a better determination of the bandpass. 
In order to reduce the data volume, the LL and RR polarisations were combined as Stokes I  and every $5$ channels were combined to  form $18$ channels of width $0.3125$\,MHz yielding an effective bandwidth of $5.625$~MHz. After this calibration, the $uv$-data from all four days for each target, were combined.

Initial imaging of each target field was done after a phase-only calibration against a model field constructed from NVSS sources within each field. 
Table~\ref{tab:cleaning} lists the important imaging parameters. The calibration was then improved  by several rounds of phase-only self-calibration followed by one round of amplitude and phase self-calibration where gain solutions were determined on a longer time-scale than the phase-only solutions. Excessive visibilities were determined from the model-subtracted data and were removed. Additional automated removal of bad data causing ripples in the image plane was done by Fourier transforming the model-subtracted images and identifying and removing extraneous peaks in the $uv$-plane. 
Further, persistent RFI was flagged and low level RFI modelled and subtracted using the LowFRFI\footnote{Obit Development Memo Series \# 16 see \url{http://www.cv.nrao.edu/~bcotton/Obit.html}} routine in ObitTalk \citep{Cotton2008}.  After self-calibration the $rms$ noise in the inner half of the primary beam area was $2.5$~mJy~beam$^{-1}$ in the central field and $3.5-5$~mJy~beam$^{-1}$ in the flanking fields, with the local noise increasing  $2-3$ times near the brightest sources. Note, the presence of extremely bright sources {with peak flux densities of the order of $5-8$~Jy~beam$^{-1}$ prior to primary beam correction} in flanking fields D through F resulted in the slightly higher overall noise in these fields.

\begin{table}
 \begin{center}
 \caption{Final \textsc{Clean}ing (top) and SPAM (bottom) parameters for individual Bo\"{o}tes fields.}
 \label{tab:cleaning}
\begin{tabular}{lc}
\hline 
Parameter & Value \\
\hline
Widefield imaging  & polyhedron facet-based\tablefootmark{a}  \\
                   & multi-frequency synthesis\tablefootmark{b}  \\
Deconvolution & Cotton-Schwab \textsc{Clean}\tablefootmark{c} \\
field size & $4\degr$ \\
facets & $85$ \\
facet size & $32.4\arcmin$\\
facet separation & $26.4\arcmin$\\
Weighting & Robust\tablefootmark{e} $-0.5$\tablefootmark{d} \\
\textsc{uvbxfn}, \textsc{uvbox} & $4$, $1$ \\
\textsc{Clean} box threshold & $5\,\sigma$ \\
\textsc{Clean} depth & $3\,\sigma$ \\
{Pixel size} & $3.8\arcsec$ \\
Restoring beam & $25\arcsec$ circular\tablefootmark{d}\\
\hline
\textsc{Spam} calibration cycles & 3 \\
Peeled sources & 20\tablefootmark{f}\\
Layer heights (weights) & $250$~km ($0.5$) \\
                        & $350$~km ($0.5$) \\
Turbulence parameter $\gamma$ & $5/3$\tablefootmark{g} \\
Model parameters & $\leq 20$\\
Reference catalogue & NVSS\tablefootmark{h} \\
\hline
\multicolumn{2}{l}{\tablefoottext{a}{\cite{Perley1989,CornwellPerley1992} } }\\
\multicolumn{2}{l}{\tablefoottext{b}{\cite{Conway+1990} } }\\
\multicolumn{2}{l}{\tablefoottext{c}{\cite{Schwab1984,Cotton1999,Cornwell+1999} } }\\
\multicolumn{2}{l}{\tablefoottext{d}{Final imaging parameters}} \\
\multicolumn{2}{l}{\tablefoottext{e}{\cite{Briggs1995}}} \\
\multicolumn{2}{l}{\tablefoottext{f}{14 for field F }}\\
\multicolumn{2}{l}{\tablefoottext{g}{pure Kolmogorov turbulence}} \\
\multicolumn{2}{l}{\tablefoottext{h}{\cite{Condon+1994,Condon+1998}}} \\
 \end{tabular}
 \end{center}
\end{table}

\begin{figure}
\resizebox{\hsize}{!}{\includegraphics{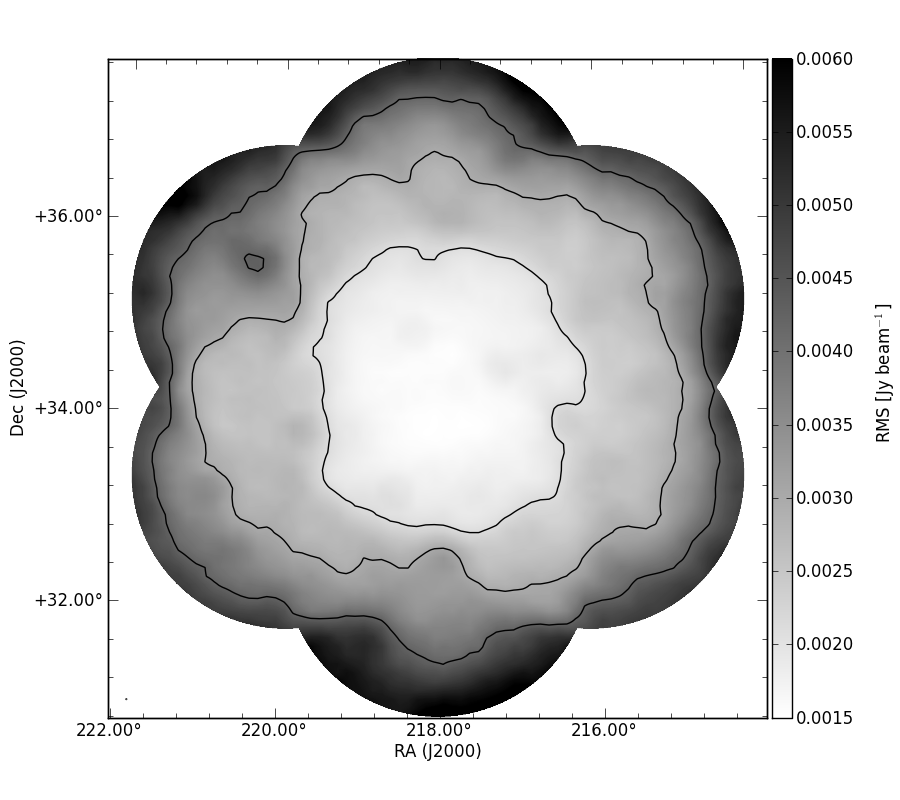}}
\caption{Greyscale map showing the local $rms$ noise measured in the mosaic image. The greyscale shows the $rms$ noise from $0.5\sigma_{avg}$ to $2\sigma_{avg}$, where $\sigma_{avg} = 3.0$~mJy~beam$^{-1}$ is the approximate $rms$ in the mosaic centre. The contours are plotted at $[1/\sqrt{2}, 1, \sqrt{2}] \times \sigma_{avg}$.  Peaks in the local noise coincide with the locations of bright sources.}
\label{fig:mosrms}
\end{figure}

Significant artefacts, however, remained in all fields near bright sources. To reduce these we applied the SPAM algorithm on the self-calibrated data. The SPAM parameters are listed in the bottom part of Table~\ref{tab:cleaning} {which include the number  of ionospheric layers modeled and their heights and relative weights,  the slope of the assumed power law function of phase structure resulting from turbulence ($\gamma$) and the number of free parameters in the fit} ; see \cite{Intema+2009} for a more detailed description of the meaning of these parameters. Three iterations of peeling were done: in the first we only applied the peeling solutions to the peeled sources and in the final two we fitted an ionospheric phase screen to the  peeling solutions. Up to $20$ sources {with flux densities above $0.4$~Jy (not corrected for primary beam effects)} were peeled in the final stage in each field. The screen was made up of two equally-weighted turbulent layers at $250$ and $350$~km. SPAM also allowed for the determination of and correction for antenna-based phase discontinuities.

\begin{figure*}
 \centering
\resizebox{\hsize}{!}{\includegraphics{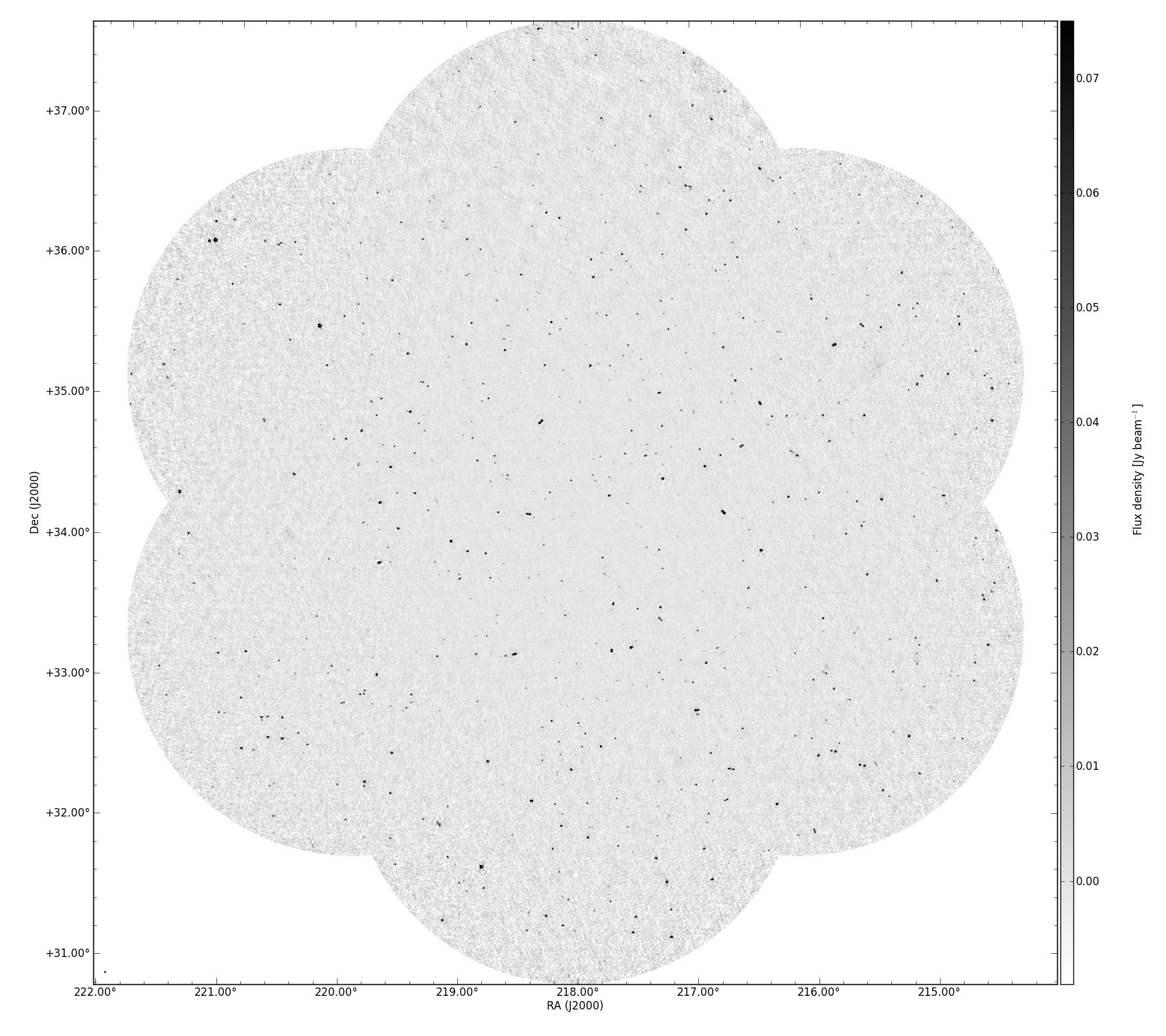}}
\caption{Greyscale map showing the entire mosaic. The image covers $30$ square degrees. The greyscale shows the flux density from $-3\sigma_{avg}$ to $25\sigma_{avg}$ where  $\sigma_{avg} = 3.0$~mJy~beam$^{-1}$ is the average $rms$ across the entire mosaic.}
\label{fig:mosaic}
\end{figure*}

\begin{figure*}
 \centering
\resizebox{\hsize}{!}{\includegraphics{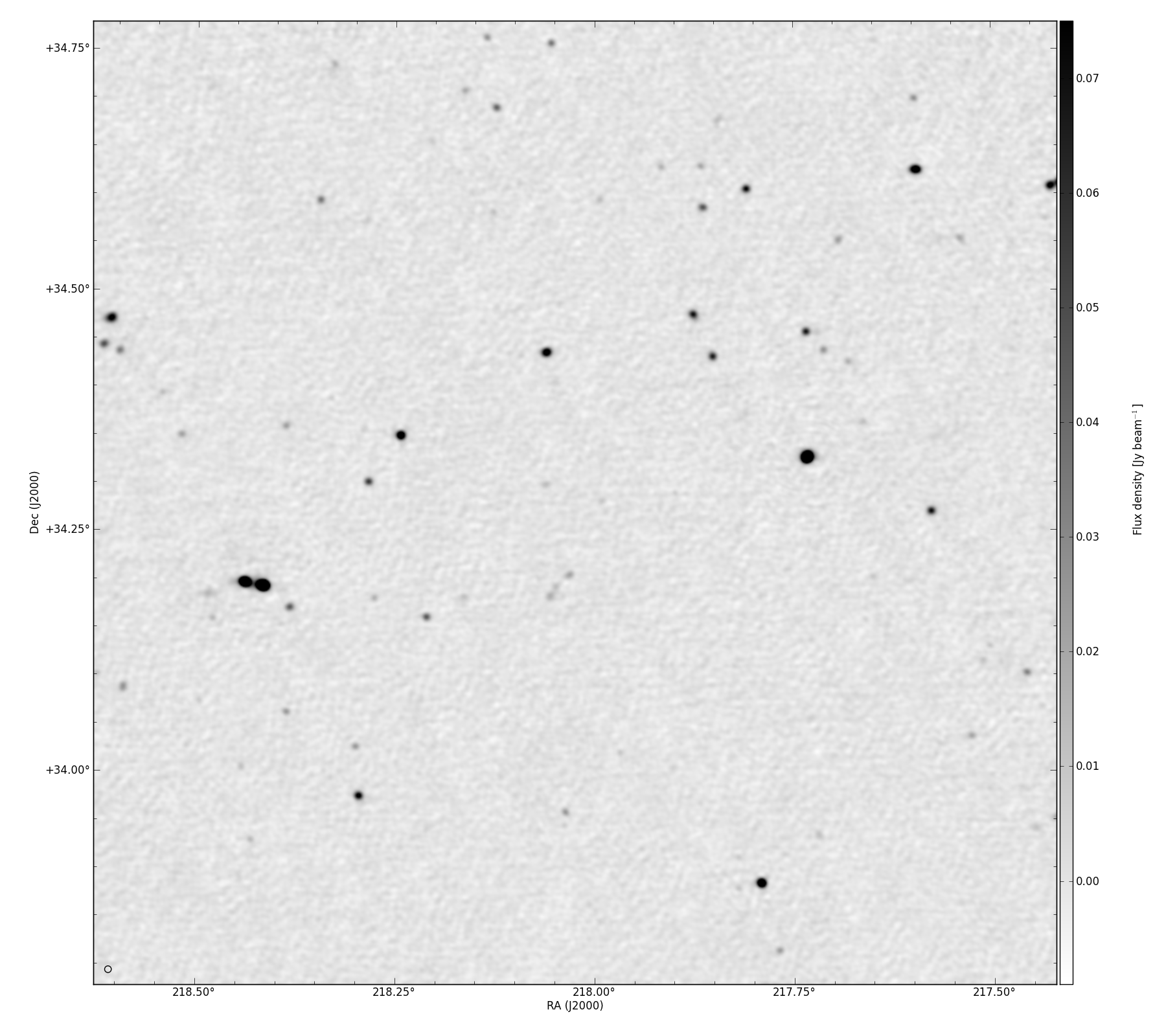}}
\caption{Zoom-in of the central part of the mosaic. The image covers $1$ square degree. The greyscale shows the flux density from  $-3\sigma_{avg}$ to $25\sigma_{avg}$ where  $\sigma_{avg} = 3.0$~mJy~beam$^{-1}$ is the average $rms$ across the entire mosaic.}
\label{fig:zoominnice}
\end{figure*}

In order to have a homogeneous point spread functions in all pointings, final images were made with a circular restoring beam of radius $25\arcsec$ {and a pixel size of $3.8\arcsec$}. {The flux density scales of the central and flanking fields were scaled up by $60$~per~cent  and $30$~per~cent respectively}  based on information from 3C\,286  (discussed in Sect. \ref{sect:errors}). In the final individual field images, the $rms$ noise in the central half of the primary beam area before primary beam correction was $1.8$~mJy~beam$^{-1}$ in the central field and $2.5 - 2.7$~mJy~beam$^{-1}$ respectively in the flanking fields. This is $3 - 5$ times the theoretical noise, similar to the factor above thermal noise obtained by deeper single pointing of \cite{Intema+2011}. The seven pointings were each corrected for the primary beam of the GMRT {up to a radius of $1.6\degr$, where the primary beam correction factor drops to $40$ per cent of its central value,} and were then mosaicked together by weighting the final image by the inverse of the square of the $rms$ noise of each individual pointing. Figure~\ref{fig:mosrms} illustrates the variation in $rms$ noise across the mosaic which is shown in entirety in  Fig.~\ref{fig:mosaic}. The noise level is smooth and around $2$~mJy~beam$^{-1}$ across the interior of the map, and increases towards the edges to about $4-5$~mJy~beam$^{-1}$. The average noise in the final mosaic is $3.0$~mJy~beam$^{-1}$, with $49$ per cent under $3$~mJy~beam$^{-1}$ and $74$ per cent under $4$~mJy~beam$^{-1}$. A small portion of the mosaic covering the inner square degree is shown in Fig.~\ref{fig:zoominnice} to illustrate the resolution and quality of the map. There remain some phase artefacts visible around the brightest sources, which have not been entirely removed during peeling. It is possible that some artefacts are caused by elevation-dependent pointing errors, since  each pointing was observed in a series of scans with varying elevations \citep{Tasse+2007,Mohan+2001,Chandra+2004}.

%

\section{Source Detection and Characterisation}
\label{sect:sources}
\subsection{Detection}
\label{sect:sources_detect}
We used the PyBDSM package\footnote{\url{http://home.strw.leidenuniv.nl/~mohan/anaamika}} to detect and characterise sources in the mosaic image. PyBDSM identifies islands of contiguous emission by identifying all pixels greater than the pixel threshold and adding each of these pixels to an island of  contiguous pixels exceeding the island threshold. Each island is fit with one or more Gaussians which are subsequently grouped into sources. Sources are classified as `S' for single sources, `M' for multiple-Gaussian sources and `C' for components of a multi-source island. From the fitted parameters the deconvolved sizes are computed assuming the theoretical beam. Errors on the fitted parameters are computed following \cite{Condon1997}.  Prior to source detection the local background $rms$ is determined by measuring the pixel statistics within a sliding box. For determining the rms background in our map we used a box size of $100$\ pixels to capture the variation in local noise around the brightest sources. We used a pixel threshold of $5\sigma_L$ and an island threshold of $3\sigma_L$. In generating a source list we allowed all Gaussians in each island to be grouped into a single source. PyBDSM detected $1296$ sources from $1578$ Gaussians fitted to $1301$ islands, of which $1073$ were single-component `S' sources. Based on visual inspection {a small number of sources  were removed} as they were {false, or bad,} detections on the edge of the image. 

The final catalogue consists of $1289$ sources between $4.1$\ mJy and $7.3$\ Jy and is available as part of the online version of this article and from the CDS\footnote{\url{http://cdsweb.u-strasborg.fr/}}. {The flux scales of the individual pointings were adjusted prior to mosaicing as described in Sect.~\ref{sect:errors} and the astrometry in the catalogue has been corrected for a systematic offset also described in Sect.~\ref{sect:errors}}. A sample of the catalogue is shown in Table~\ref{tab:cataloguesample} where the columns are: (1) Source name, (2,3) flux-weighted position right ascension, RA, and {uncertainty}, (4,5) flux-weighted position declination, DEC, and {uncertainty}, (6) integrated source flux density and uncertainty, (7) peak flux density and uncertainty, (8,9,10) fitted parameters: deconvolved major- and minor-axes, and position angle, for extended sources, (11) local $rms$ noise, and (12) the number of Gaussians fitted to the source. Extended sources are classified as such based on the ratio between the integrated and peak flux densities (see Sect.~\ref{sect:sources_ext}). Unresolved sources have a `-' listed for all their fitted shape parameters (semi-major and -minor axes and position angle) or for only the semi-minor axis where the source is resolved in one direction. For extended sources consisting of multiple Gaussians, the fitted parameters for each Gaussian are given on separate lines in the table, listed as `a', `b', etc.  Images of the $25$ brightest sources are shown in Appendix~\ref{app:images}.
\begin{table*}
\tiny
 \centering
 \begin{center}
 \caption{Sample of the GMRT $153$\ MHz source and Gaussian-component catalogue.}
 \label{tab:cataloguesample}
\begin{tabular}{rccccr@{\,$\pm$\,}lr@{\,$\pm$\,}lr@{\,$\pm$\,}lr@{\,$\pm$\,}lr@{\,$\pm$\,}lcc}
\hline
\multicolumn{1}{c}{Source ID} & \multicolumn{1}{c}{RA} & \multicolumn{1}{c}{$\sigma_{RA}$} & \multicolumn{1}{c}{DEC}  & \multicolumn{1}{c}{$\sigma_{DEC}$} & \multicolumn{2}{c}{$S_i$}  & \multicolumn{2}{c}{$S_p$}  & \multicolumn{2}{c}{$a$\tablefootmark{a}} & \multicolumn{2}{c}{$b$\tablefootmark{a}} & \multicolumn{2}{c}{$\phi$\tablefootmark{a}} & \multicolumn{1}{c}{$rms$}  & \multicolumn{1}{c}{$N_{gauss}$\tablefootmark{b}}\\
  & \multicolumn{1}{c}{[deg]} & \multicolumn{1}{c}{[$\arcsec$]} & \multicolumn{1}{c}{[deg]} & \multicolumn{1}{c}{[$\arcsec$]} & \multicolumn{2}{c}{[mJy]} & \multicolumn{2}{c}{[mJy\ beam$^{-1}$]} & \multicolumn{2}{c}{[$\arcsec$]} & \multicolumn{2}{c}{[$\arcsec$]} & \multicolumn{2}{c}{[deg]} & \multicolumn{1}{c}{[mJy\ beam$^{-1}$]} &  \\
\multicolumn{1}{c}{(1)} & \multicolumn{1}{c}{(2)} & \multicolumn{1}{c}{(3)} & \multicolumn{1}{c}{(4)} & \multicolumn{1}{c}{(5)} & \multicolumn{2}{c}{(6)} & \multicolumn{2}{c}{(7)} & \multicolumn{2}{c}{(8)} & \multicolumn{2}{c}{(9)} & \multicolumn{2}{c}{(10)} & \multicolumn{1}{c}{(11)} & \multicolumn{1}{c}{(12)} \\
\hline
J144733+3507 & 221.88796 & 0.9 & 35.13126 & 1.5 & 158 & 33 & 69 & 15 & 38.3 & 3.4 & 19.3 & 1.9 & 5 & 6 & 5.4 & 1 \\
J144658+3308 & 221.74358 & 2.9 & 33.14058 & 1.1 & 42 & 13 & 20 & 6 & \multicolumn{2}{c}{ } & \multicolumn{2}{c}{ } & \multicolumn{2}{c}{ } & 4.9  & 2 \\
 a & 221.74843 & 2.8 & 33.14096 & 1.8 & 19 &  8 & 19 & 5 & \multicolumn{2}{c}{-} & \multicolumn{2}{c}{-} & \multicolumn{2}{c}{-} &    &  \\
 b & 221.73914 & 2.8 & 33.14001 & 2.6 & 23 &  8 & 18 & 5 & 15.9 & 6.8 & 9.5 & 5.6 & 56 & 90 &    &  \\
J144705+3442 & 221.77266 & 1.8 & 34.71605 & 1.6 & 32 &  9 & 27 & 7 & \multicolumn{2}{c}{-} & \multicolumn{2}{c}{-} & \multicolumn{2}{c}{-} & 4.7  & 1 \\
J144706+3457 & 221.77552 & 1.4 & 34.95096 & 1.5 & 25 &  9 & 28 & 7 & \multicolumn{2}{c}{-} & \multicolumn{2}{c}{-} & \multicolumn{2}{c}{-} & 5.0  & 1 \\
J144645+3330 & 221.69153 & 1.8 & 33.50822 & 1.4 & 31 &  9 & 29 & 7 & \multicolumn{2}{c}{-} & \multicolumn{2}{c}{-} & \multicolumn{2}{c}{-} & 4.8  & 1 \\
J144640+3322 & 221.66919 & 0.9 & 33.36920 & 0.8 & 72 & 16 & 62 & 13 & \multicolumn{2}{c}{-} & \multicolumn{2}{c}{-} & \multicolumn{2}{c}{-} & 4.9  & 1 \\
J144646+3440 & 221.69543 & 1.5 & 34.67482 & 1.1 & 29 &  9 & 32 & 8 & \multicolumn{2}{c}{-} & \multicolumn{2}{c}{-} & \multicolumn{2}{c}{-} & 5.1  & 1 \\
J144648+3546 & 221.70230 & 1.9 & 35.77421 & 1.4 & 19 &  8 & 25 & 6 & \multicolumn{2}{c}{-} & \multicolumn{2}{c}{-} & \multicolumn{2}{c}{-} & 5.0  & 1 \\
J144613+3303 & 221.55831 & 0.7 & 33.06494 & 0.6 & 169 & 35 & 131 & 27 & \multicolumn{2}{c}{-} & \multicolumn{2}{c}{-} & \multicolumn{2}{c}{-} & 5.0  & - \\
J144638+3553 & 221.65996 & 2.1 & 35.88845 & 2.3 & 37 & 11 & 27 & 7 & \multicolumn{2}{c}{-} & \multicolumn{2}{c}{-} & \multicolumn{2}{c}{-} & 5.7  & 1 \\
J144626+3512 & 221.61124 & 1.0 & 35.20927 & 0.7 & 314 & 65 & 244 & 50 & \multicolumn{2}{c}{-} & \multicolumn{2}{c}{-} & \multicolumn{2}{c}{-} & 7.5  & - \\
J144619+3425 & 221.58123 & 3.2 & 34.42447 & 2.4 & 46 & 12 & 26 & 7 & \multicolumn{2}{c}{-} & \multicolumn{2}{c}{-} & \multicolumn{2}{c}{-} & 5.5  & 1 \\
J144606+3316 & 221.52865 & 2.9 & 33.26922 & 2.2 & 20 &  7 & 17 & 5 & \multicolumn{2}{c}{-} & \multicolumn{2}{c}{-} & \multicolumn{2}{c}{-} & 4.2  & 1 \\
J144557+3251 & 221.49062 & 0.6 & 32.86032 & 0.5 & 122 & 26 & 111 & 23 & \multicolumn{2}{c}{-} & \multicolumn{2}{c}{-} & \multicolumn{2}{c}{-} & 5.0  & 1 \\
J144555+3237 & 221.48098 & 2.4 & 32.62328 & 3.6 & 24 &  7 & 17 & 5 & \multicolumn{2}{c}{-} & \multicolumn{2}{c}{-} & \multicolumn{2}{c}{-} & 4.4  & 1 \\
J144617+3506 & 221.57405 & 1.1 & 35.11639 & 1.3 & 158 & 34 & 75 & 16 & \multicolumn{2}{c}{ } & \multicolumn{2}{c}{ } & \multicolumn{2}{c}{ } & 5.3  & 2 \\
 a & 221.57091 & 0.9 & 35.11314 & 0.7 & 93 & 20 & 78 & 16 & 15.3 & 1.8 & 5.6 & 1.4 & 72 & 16 &    &  \\
 b & 221.57923 & 1.4 & 35.12223 & 2.1 & 65 & 15 & 38 & 9 & 29.4 & 5.1 & 10.6 & 2.8 & 12 & 14 &    &  \\
J144602+3339 & 221.50949 & 2.1 & 33.66347 & 1.3 & 42 & 10 & 27 & 6 & 29.3 & 5.3 & \multicolumn{2}{c}{-} & 64 & 11 & 3.8 & 1 \\
J144607+3503 & 221.52967 & 1.2 & 35.05975 & 0.6 & 81 & 18 & 62 & 13 & \multicolumn{2}{c}{-} & \multicolumn{2}{c}{-} & 178 & 6 & 4.8 & 1 \\
\hline
\multicolumn{17}{l}{\tablefoottext{a}{Parameters are given for extended sources to which Gaussian components were successfully fit.} }\\
\multicolumn{17}{l}{\tablefoottext{b}{A ``-'' indicates a poor Gaussian fit. In these cases the total flux density quoted is the total flux density in the source island.} }\\
 \end{tabular}
 \end{center}
\end{table*}

\subsection{Resolved Sources}
\label{sect:sources_ext}
\begin{figure*}
 \centering
\resizebox{0.48\hsize}{!}{\includegraphics{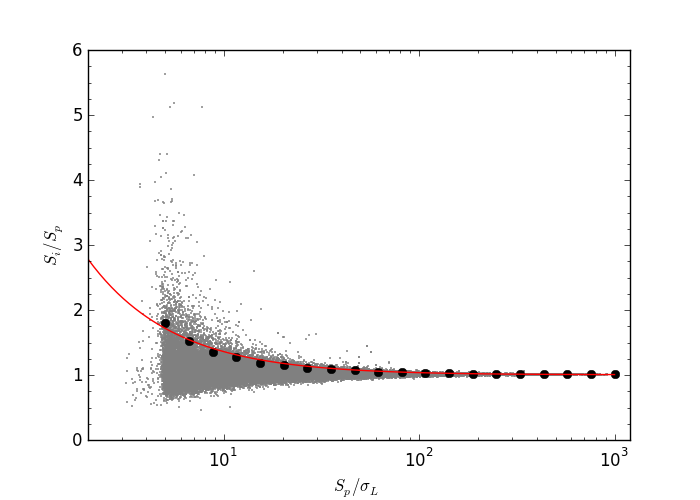}}
\resizebox{0.48\hsize}{!}{\includegraphics{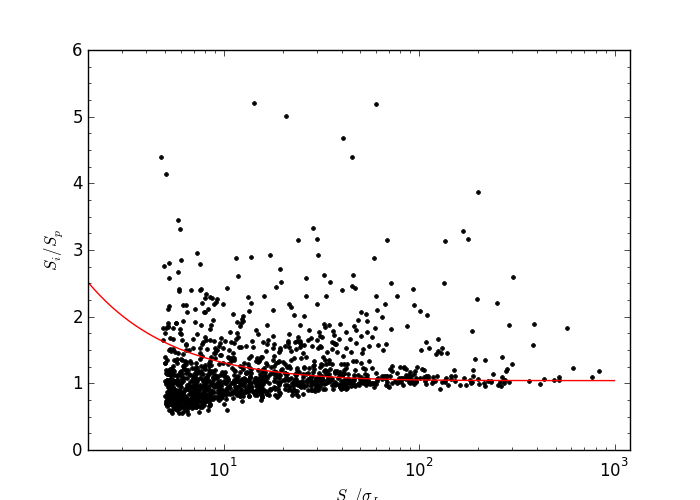}}
\caption{\textit{Left} The simulated ratio of integrated to peak flux density as a function of signal-to-noise ratio for sources from the $25$ Monte-Carlo simulations. For $20$ logarithmic bins in signal-to-noise ratio, the black points show the threshold below which 95~per~cent of the sources lie in that bin.   The red line shows a fit to this upper envelope. \textit{Right} The measured ratio of integrated to peak flux density as a function of signal-to-noise ratio. The line shows the upper envelope containing 95~per~cent of the unresolved sources as determined from Monte-Carlo simulations. }
\label{fig:resolvedsim}
\end{figure*}

In the presence of no noise, the extendedness of a source can simply be determined from the ratio of the integrated flux density to the peak flux density, $S_i/S_p > 1$. However, since the errors on $S_i$ and $S_p$ are  correlated, the $S_i/S_p$ distribution is skewed, particularly at low signal-to-noise. To determine an upper envelope of this distribution, we performed a Monte-Carlo simulation in which we generated $25$ random fields containing $\sim\!10\,000$~randomly positioned point sources with peak flux densities between $0.1\sigma$ and $20\sigma$, where $\sigma$ was taken to be $3$~mJy~beam$^{-1}$. The source flux densities are drawn randomly from the source count distribution, $dN/dS \propto S^{-1.6}$ (Sect.~\ref{sect:source_counts}). We neglect the deviation of the true source counts from a power law slope at high fluxes as there are very few sources at these fluxes. The $rms$ noise map for these fields was taken from the central $4000\times4000$ pixel$^2$ of the residual mosaic. Source detection was performed in the same manner described in Sect.~\ref{sect:sources_detect}, thus only $\sim\!750$ sources in each field satisfy the detection criterion of peak flux density $>5\sigma$. The $S_i/S_p$ distribution produced from the Monte-Carlo simulation is plotted in {the \textit{left} panel of Fig.~\ref{fig:resolvedsim}}. {To determine the $95$~per~cent envelope, a curve was fit to the $95$th percentile of $20$ logarithmic bins across signal-to-noise ratio. The fitted envelope is characterised by:}
{\[ S_i/S_p = 1+ \left\lbrace (0.01\pm0.02)^2 + (3.58\pm0.10 )^2 \left(\sigma_L/S_p\right)^2 \right\rbrace^{0.5}. \]} The measured distribution of $S_i/S_p$ as a function of signal-to-noise ratio is shown in the \textit{right} panel of Fig.~\ref{fig:resolvedsim}. The line shows the upper envelope from the Monte-Carlo simulation. {Of the $453$ sources that lie above this line ($35$~per~cent of all $1289$ sources), approximately $41$, i.e. $9$~per~cent, are not truly extended and merely lie above the line by chance. However, all these sources above the line are listed in the catalogue as extended and the measured deconvolved FWHM major and minor axes are given. }

\subsection{Completeness and Reliability}
To quantify the completeness and reliability of the catalogue, we performed a similar Monte-Carlo simulation to that described in the previous section. However, in this case approximately $25$~per cent of the artificial sources inserted into the noise map were extended sources -- Gaussians with FWHM larger than the beamsize. This allows for a better estimate of the completeness and reliability in terms of integrated flux densities.

\begin{figure*}
 \centering
\resizebox{0.48\hsize}{!}{\includegraphics{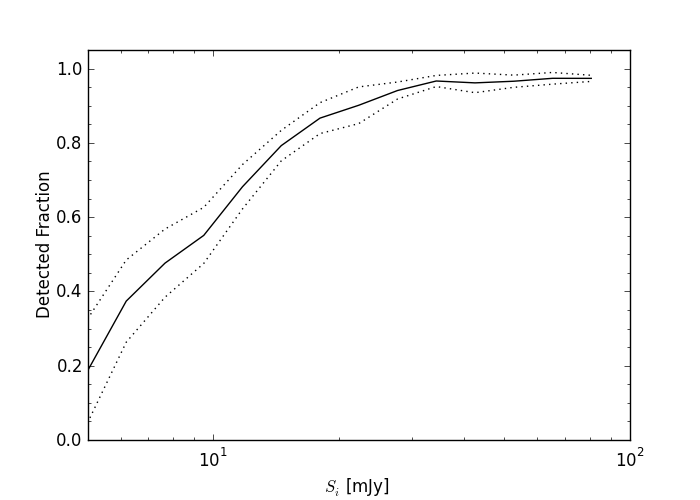}}
\resizebox{0.48\hsize}{!}{\includegraphics{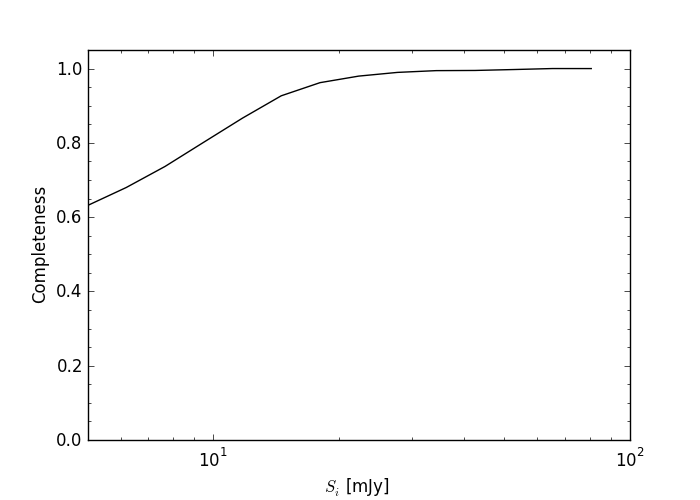}}
\caption{\textit{Left} Fraction of sources detected as a function of integrated flux density to local noise ratio calculated from $25$ Monte-Carlo simulations. The solid line shows the mean of all $25$ randomly generated fields and the two dotted lines show the $1\sigma$ uncertainty. \textit{Right} Estimated completeness of the catalogue as a function of integrated flux density limit {accounting for the varying sensitivity across the field of view}. }
\label{fig:complete}
\end{figure*}

The completeness of a catalogue represents the probability that all sources above a given flux density are detected. We have estimated this by plotting the fraction of detected sources{ in our MC simulation} as a function of integrated flux density {(\textit{left} panel of Fig.~\ref{fig:complete})}, i.e. the fraction of input sources that have a catalogued flux density {using the same detection parameters}. Due to the variation in $rms$ across the image, the detection fraction has been multiplied by the fraction of the total $30$\ deg$^2$ area in which the source can be detected. The completeness at a given flux density is determined by integrating the detected fraction upwards from a given flux density limit and is plotted as a function of integrated flux density in {the \textit{right} panel of} Fig.~\ref{fig:complete}. We thus estimate that the catalogue is $95$~per~cent complete above a peak flux density of $14$~mJy.

\begin{figure*}
 \centering
\resizebox{0.48\hsize}{!}{\includegraphics{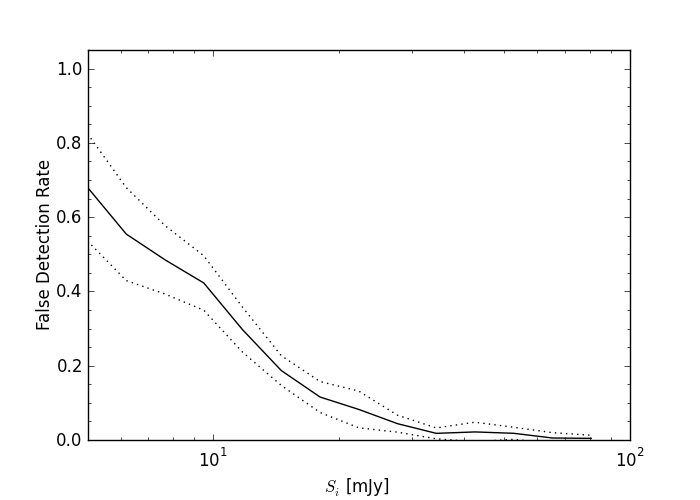}}
\resizebox{0.48\hsize}{!}{\includegraphics{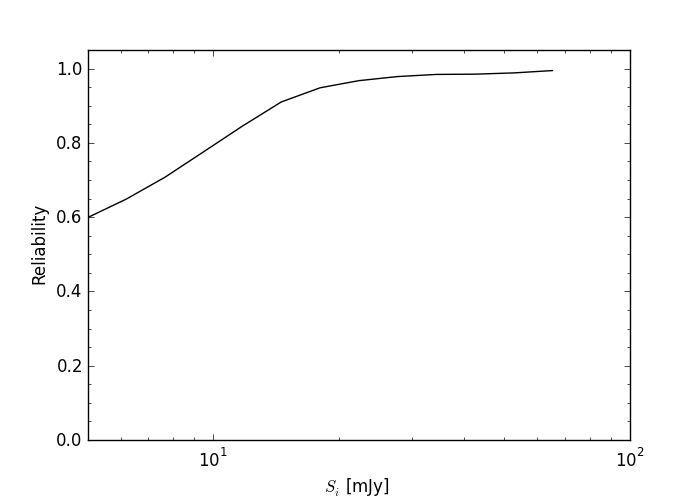}}
\caption{{\textit{Left} False detection rate as a function of peak flux density to local signal-to-noise ratio calculated from $25$ Monte-Carlo simulations. The solid line shows the mean of all $25$ randomly generated fields and the two dotted lines show the $1\sigma$ uncertainty. \textit{Right} Estimated reliability of the catalogue as a function of integrated flux density limit accounting for the varying sensitivity across the field of view.} }
\label{fig:rely}
\end{figure*}

The reliability of the catalogue indicates the probability that all sources above a given flux density are real. In {the \textit{left} panel of} Fig.~\ref{fig:rely},  the false detection rate $FDR$, i.e. the fraction of catalogued sources that do not have an input source, is plotted as a function of the integrated flux density. Integrating up from a  given detection limit and multiplying by the normalised source flux distribution, we can determine an estimate of the overall FDR or reliability, $R = 1-FDR$, of the catalogue. {The reliability is plotted as a function of integrated flux density limit in the \textit{right} panel of Fig.~\ref{fig:rely}.} For a $14$~mJy detection threshold, the reliability is ${92}$~per~cent.

\subsection{Astrometric and Flux Uncertainties}
\label{sect:errors}
Errors in the phase calibration introduce uncertainties in the source positions. To assess these uncertainties and determine any systematic offsets we selected a sample of sources with peak flux densities at least $10\sigma_L$. We searched for $1.4$~GHz NVSS \citep{Condon+1998} sources within $45\arcsec$ of these targets. $745$ matches were found. From this sample, we measured a small offset of $(\Delta \alpha, \Delta \delta) = (0.44\arcsec, -0.21\arcsec)$, which is of the order of the pixel size of the $153$~MHz observations and the NVSS accuracy ($\sim\!1\arcsec$). A correction for this offset has been applied to all sources in the catalogue. The scatter in the offsets between the GMRT and NVSS positions is a combination of noise-independent  calibration errors, $\epsilon$, in both the GMRT and NVSS data as well as a noise-dependent error, $\sigma$, from position determination via Gaussian-fitting:
\[ \sigma^2 = \epsilon_{GMRT}^2 + \epsilon_{NVSS}^2  + \sigma_{GMRT}^2 + \sigma_{NVSS}^2\]
From \cite{Condon+1998}, the NVSS calibration errors are $(\epsilon_{\alpha}, \epsilon_{\delta})_{NVSS} = (0.45\arcsec, 0.56\arcsec)$. To separate the noise-dependent and -independent uncertainties we select from the above sample only the NVSS sources with position errors of less than $0.6\arcsec$ and measure a scatter of $(\sigma_{\alpha}, \sigma_{\delta})_{GMRT} = (0.67\arcsec, 0.65\arcsec)$.  For this very high signal-to-noise sub-sample of $107$ sources the noise-dependent fit errors for both the GMRT and NVSS can safely be assumed to be small so we determine the GMRT calibration errors to be $(\epsilon_{\alpha}, \epsilon_{\delta})_{GMRT} = (0.50\arcsec, 0.32\arcsec)$. These are added quadratically to the Gaussian-fit position uncertainties in the catalogue.


Similarly, in addition to the noise-dependent Gaussian fitting uncertainties on the fluxes \citep{Condon1997}, the uncertainty in the measured flux densities all consists of a noise-independent  component. The uncertainty introduced  through transferring the flux density scale from the calibrator to the target fields is the main such uncertainty and depends on a number of factors: (i) the data quality, (ii) the accuracy of the model, and (iii) differences in observing conditions between the calibrator and target. 

Like the target data, the calibrator data is adversely affected by RFI and the ionosphere. The RFI conditions of the flanking field observations were similar to those when the central pointing data were taken, however, the ionosphere was not as calm. Following \cite{Intema+2011} we adopt a slightly inflated, ad-hoc amplitude uncertainty of $\sim\!4$~per~cent due to RFI and ionospheric effects. 

The calibrator model is of a point source whose flux density at $153$~MHz is predicted from the Perley-Taylor model based on flux density measurements at many frequencies. 3C\,48 is a point source of $64.4$~Jy at $153$~MHz. \cite{Intema+2011} provide an improved model for 3C\,286, a point source of $31.01$~Jy at $153$~MHz, and estimate a flux density uncertainty of $5$~per~cent. The large field of view, however, means that there are other fainter sources present in the calibrator field. For similar duration  observations of 3C\,286 \cite{Intema+2011} set an upper limit of $1$~per~cent on the flux density uncertainty. Since 3C\,48 is about a factor of two brighter, we estimate that the flux density uncertainty due to additional sources in the 3C\,48 field is also at most $1$~per~cent.

Individual antennas are sensitive to the galactic diffuse radio emission which varies across the sky and so may be different for the calibrator and target fields thereby introducing an offset to the flux density scale as well as additional uncertainty. However, since the GMRT does not measure the sky temperature, we require external information to take this into account. Following \cite{Tasse+2007} and \cite{Intema+2011} we determine the mean off-source sky-temperature from the \cite{Haslam+1982} all-sky radio maps at $408$~MHz: both the Bo\"{o}tes and 3C\,286 fields have sky temperatures of $\sim\!20 \pm 1$~K. Using the equation from \cite{Tasse+2007}, this implies that no offset in the flux density scale is required for 3C\,286 and we estimate a gain uncertainty of $2$~per cent. However, the sky temperature near the primary calibrator 3C\,48 is $24 \pm 1$~K which implies a {flux density correction of $0.92$  with an estimated uncertainty of $8$ per cent}. Since the flux density scale is linear, this offset is applied post hoc to the measured flux densities.

Prior to combining the individual pointings, we compared the measured {primary beam-corrected} flux densities of sources in the overlapping regions {(approximately $110-150$ sources per region)} and found those in the flanking fields to be consistently higher by $30 \pm 5$ per cent. To investigate this we made images after calibration using 3C\,286 as the primary calibrator. This yielded consistent fluxes between the central and flanking fields. It is likely that significant time-dependent changes in the GMRT systems over the course of each observing night were captured by the  regular (each $30$ min)  observations of 3C\,286. We thus used the 3C\,286-calibrated images to derive a correction to the flux density scales of the flanking fields, a factor of $1.3$, before combining the individual pointings. The uncertainty of this correction is $10$~per~cent.

The total estimated uncertainty in transferring the flux density scale is of the order of $20$~per~cent which we add quadratically to the measured Gaussian fit uncertainty for each source. Comparison of the flux density of bright sources measured in the individual pointings after the above correction shows good agreement between the flux density scales of the individual pointings and the measured scatter is $\sim\!16$~per~cent, which also includes a contribution by the noise-dependent terms. 

\subsection{Diffuse Sources}
\label{sect:diffuse}

\begin{figure}
 \centering
\resizebox{0.75\hsize}{!}{\includegraphics{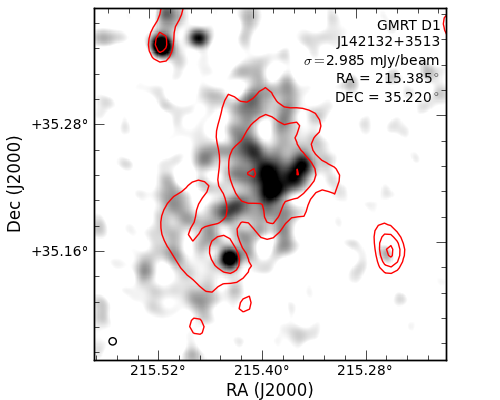}}
\resizebox{0.75\hsize}{!}{\includegraphics{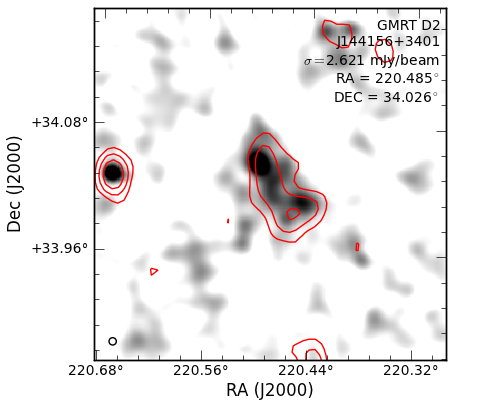}}
\caption{Postage stamps showing D1, RA=14:21:32, DEC=+35:12:12 (\textit{top}) and D2, RA=14:41:56,  DEC=+34:01:34 (\textit{bottom}). The greyscale goes from $0.5\sigma$ to $5\sigma$ and the images have been smoothed with a Gaussian of $50$~arcsec. WENSS contours are overlaid at $[1.5,3.0,10.0]\times \sigma_L$ {where $\sigma_L$ is the local \textit{rms} in the WENSS images -- $3.5$~mJy~beam$^{-1}$ and $3.7$~mJy~beam$^{-1}$ respectively for D1 and D2}.}
\label{fig:diffuse}
\end{figure}

We have identified two faint diffuse sources in the final mosaic which were not detected by PyBDSM as their peak flux densities are too low. Postage stamps of these two sources are shown in Fig. \ref{fig:diffuse}. The first, D1, is located at RA = 14:21:32, DEC = +35:12:12. This source has previously detected in WENSS by \cite{DelainRudnick2006} who have associated it with a galaxy group at $z=0.01$.
The second diffuse source, D2, is located at RA = 14:41:56,  DEC = +34:01:34.

\section{Analysis}
\label{sect:analysis}
The $1289$ sources in the catalogue provide a statistically significant sample across three orders of magnitude in flux density from $4$~mJy to $7$~Jy. In this section we present the derived $153$\ MHz source counts and spectral index distributions based on matching these sources to catalogues at $1.4$\ GHz and $320$\ MHz.

\subsection{Source Counts}
\label{sect:source_counts}
\begin{figure*}
 \centering
\resizebox{0.8\hsize}{!}{\includegraphics{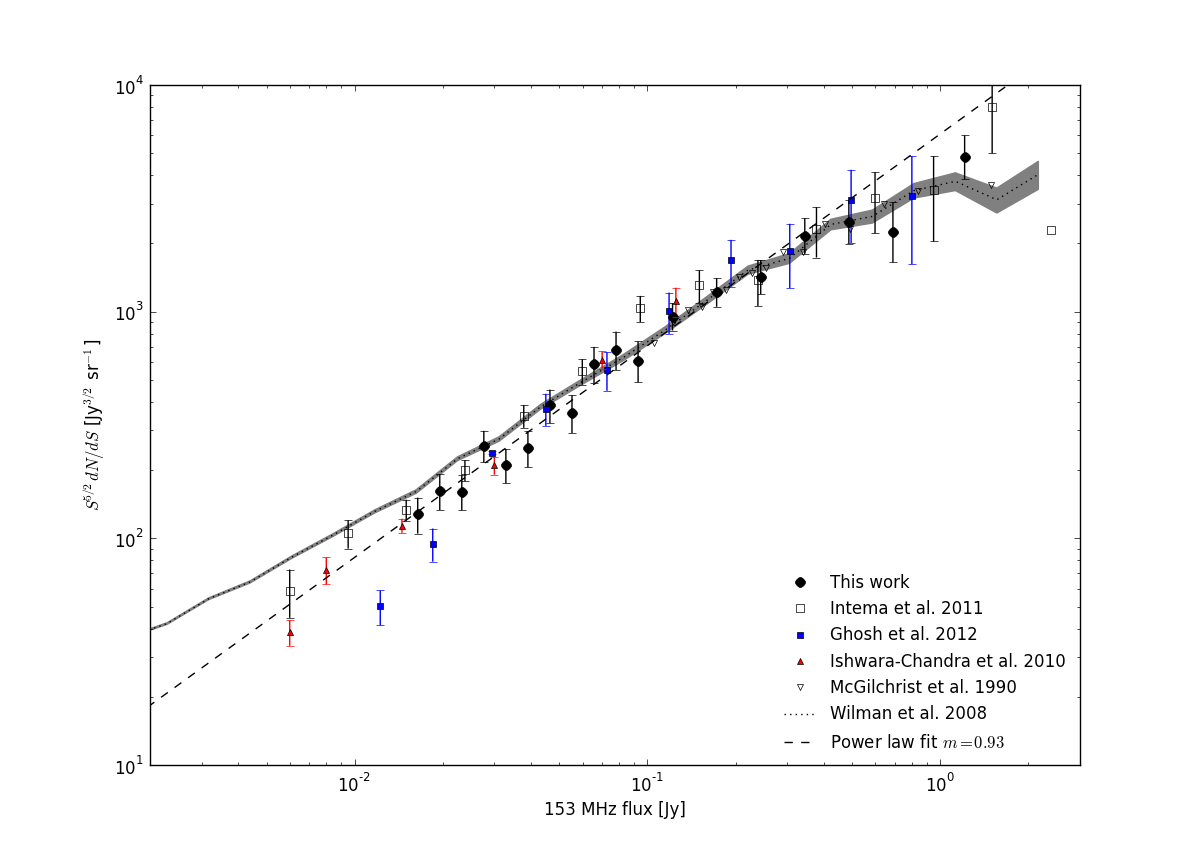}}
\caption{{Euclidean-normalized differential source counts for the GMRT $153$ MHz catalogue (filled black circles) in $18$ logarithmic flux density bins between $15$~mJy and $6.5$~Jy. For comparison we have plotted the $153$ MHz source counts from \cite{Intema+2011} (open squares) for the central B\"{o}otes pointing, from \cite{Ghosh+2012} (blue filled squares) and from \cite{IshwaraChandra+2010} (red filled triangles), as well as the $151$~MHz source counts from \cite{McGilchrist+1990} for part of the 7C catalogue (open inverted triangles). Also shown is a source count model by \cite{Wilman+2008} (dotted line with shaded area indicating the $1\sigma$ errors) and  a power law fitted between $150 - 400$~mJy (dashed line) which has a slope of  $0.93 \pm 0.04$. } }
\label{fig:srccounts}
\end{figure*}

The Euclidean-normalized differential source counts are shown in Fig.~\ref{fig:srccounts}. Due to the large variation in $rms$ across the mosaic, the sources are not uniformly detected across the image, i.e. faint sources can only be detected in a smaller area in the inner part of the image. {We therefore weight each source by the inverse of the area in which it can be detected \citep[e.g.][]{Windhorst+1985}, which also accounts for the varying detection area within a given flux density bin. Accurate derivation of the source counts is complicated by a number of effects. In general, noise can scatter sources into adjacent bins, most noticably at low flux densities.  A positive bias is introduced by the enhancement of weak sources by random noise peaks (Eddington bias). Furthermore, low surface brightness extended sources can be missed as their peak flux densities fall below the detection limit. We have used our Monte-Carlo simulations to estimate the combined contribution of these effects and derive a correction factor to the observed source counts. Errors on the final normalised source counts are propagated from the errors on the correction factors and the Poisson errors \citep{Gehrels1986} on the raw counts per bin. The flux density bins start at three times the average $rms$, $15$~mJy, and increase in factors of $2^{1/4}, 2^{1/2}$ or $2$  chosen to provide source counts of $60-80$ in most, except for the highest, flux density bins. Table~\ref{tab:src_counts} lists (i) the flux density bins, (ii) the central flux density of the bin, (iii) the raw counts, (iv) the effective detection areas for sources at the lower and upper limits of the flux density bin, (v) the effective area corresponding to the bin centre, (vi) the mean weight of the sources in the bin, (vii) the correction factor, and (viii) the corrected normalised source counts.
}

\begin{table*}
 \begin{center}
 \caption{{Euclidean-normalized differential source counts for the GMRT $153$ MHz catalogue.}}
 \label{tab:src_counts}
\begin{tabular}{cccccccc}
\hline
\multicolumn{1}{c}{$S$ Range} & \multicolumn{1}{c}{$S_c$} & \multicolumn{1}{c}{Raw Counts} & \multicolumn{1}{c}{Area} & \multicolumn{1}{c}{$A(S_c)$} & \multicolumn{1}{c}{$<W>$}&\multicolumn{1}{c}{Correction}  & \multicolumn{1}{c}{Normalised counts}\\
\multicolumn{1}{c}{[Jy]} & \multicolumn{1}{c}{[Jy]} & \multicolumn{1}{c}{ } & \multicolumn{1}{c}{[deg$^2$]} & \multicolumn{1}{c}{[deg$^2$]} & \multicolumn{1}{c}{ }&\multicolumn{1}{c}{ }  & \multicolumn{1}{c}{ [Jy$^{3/2}$ sr$^{-1}$] }\\
\hline
$0.015-0.018$ & $0.016$ & $70.0_{-8.3}^{+ 9.4}$ & $17.1-22.1$ & $19.7$ & $0.80$ & $1.14 \pm 0.15$ &   $127_{- 22}^{+  24}$ \\[+1.5pt]
$0.018-0.021$ & $0.020$ & $68.0_{-8.2}^{+ 9.3}$ & $22.1-26.5$ & $24.4$ & $0.83$ & $1.11 \pm 0.14$ &   $162_{- 28}^{+  30}$ \\[+1.5pt]
$0.021-0.025$ & $0.023$ & $66.0_{-8.1}^{+ 9.2}$ & $26.5-30.5$ & $28.7$ & $0.91$ & $1.06 \pm 0.13$ &   $160_{- 28}^{+  30}$ \\[+1.5pt]
$0.025-0.030$ & $0.028$ & $86.0_{-9.3}^{+10.3}$ & $30.5-32.9$ & $32.1$ & $0.93$ & $1.00 \pm 0.11$ &   $256_{- 40}^{+  42}$ \\[+1.5pt]
$0.030-0.036$ & $0.033$ & $62.0_{-7.9}^{+ 8.9}$ & $32.9-33.3$ & $33.2$ & $0.96$ & $0.94 \pm 0.10$ &   $210_{- 35}^{+  38}$ \\[+1.5pt]
$0.036-0.042$ & $0.039$ & $58.0_{-7.6}^{+ 8.7}$ & $\ldots$ & $33.3$ & $0.95$ & $0.90 \pm 0.11$ &   $249_{- 44}^{+  47}$ \\[+1.5pt]
$0.042-0.050$ & $0.046$ & $74.0_{-8.6}^{+ 9.6}$ & $\ldots$ & $33.3$ & $0.99$ & $0.91 \pm 0.11$ &   $386_{- 64}^{+  68}$ \\[+1.5pt]
$0.050-0.060$ & $0.055$ & $51.0_{-7.1}^{+ 8.2}$ & $\ldots$ & $33.3$ & $0.99$ & $0.93 \pm 0.11$ &   $357_{- 65}^{+  71}$ \\[+1.5pt]
$0.060-0.071$ & $0.066$ & $64.0_{-8.0}^{+ 9.0}$ & $\ldots$ & $33.3$ & $0.99$ & $0.95 \pm 0.12$ &   $588_{-105}^{+ 112}$ \\[+1.5pt]
$0.071-0.085$ & $0.078$ & $57.0_{-7.5}^{+ 8.6}$ & $\ldots$ & $33.3$ & $1.00$ & $0.96 \pm 0.13$ &   $680_{-127}^{+ 137}$ \\[+1.5pt]
$0.085-0.101$ & $0.093$ & $39.0_{-6.2}^{+ 7.3}$ & $\ldots$ & $33.3$ & $1.00$ & $0.96 \pm 0.11$ &   $606_{-119}^{+ 133}$ \\[+1.5pt]
$0.101-0.143$ & $0.122$ & $81.0_{-9.0}^{+10.0}$ & $\ldots$ & $33.3$ & $1.00$ & $0.96 \pm 0.07$ &   $951_{-126}^{+ 136}$ \\[+1.5pt]
$0.143-0.202$ & $0.172$ & $61.0_{-7.8}^{+ 8.9}$ & $\ldots$ & $33.3$ & $1.00$ & $0.97 \pm 0.05$ &   $1218_{-167}^{+ 187}$ \\[+1.5pt]
$0.202-0.285$ & $0.244$ & $42.0_{-6.5}^{+ 7.5}$ & $\ldots$ & $33.3$ & $1.00$ & $0.98 \pm 0.05$ &   $1423_{-229}^{+ 265}$ \\[+1.5pt]
$0.285-0.404$ & $0.345$ & $38.0_{-6.1}^{+ 7.2}$ & $\ldots$ & $33.3$ & $1.00$ & $0.98 \pm 0.04$ &   $2160_{-362}^{+ 422}$ \\[+1.5pt]
$0.404-0.571$ & $0.487$ & $26.0_{-5.1}^{+ 6.2}$ & $\ldots$ & $33.3$ & $1.00$ & $0.98 \pm 0.05$ &   $2496_{-502}^{+ 606}$ \\[+1.5pt]
$0.571-0.807$ & $0.689$ & $14.0_{-3.7}^{+ 4.8}$ & $\ldots$ & $33.3$ & $1.00$ & $0.98 \pm 0.05$ &   $2257_{-608}^{+ 790}$ \\[+1.5pt]
$0.807-1.615$ & $1.211$ & $25.0_{-5.0}^{+ 6.1}$ & $\ldots$ & $33.3$ & $1.00$ & $0.98 \pm 0.04$ &   $4807_{-971}^{+1182}$ \\[+1.5pt]
$1.615-6.458$ & $4.036$ & $9.0_{-2.9}^{+ 4.1}$ & $\ldots$ & $33.3$ & $1.00$ & $0.99 \pm 0.01$ &   $5925_{-1936}^{+2715}$ \\[+1.5pt]
\hline
 \end{tabular}
 \end{center}
\end{table*}

We have compared our source counts with the little observational data available at this frequency. Our source counts agree well with those derived by \cite{Intema+2011} for the central field. {Since their image is approximately three times deeper than our mosaic, the good agreement at low flux densities lends credance to our correction factors. The recent source counts from \cite{Ghosh+2012} and those by \cite{IshwaraChandra+2010} for a smaller, slightly shallower GMRT field also agree well with our data, except the \cite{Ghosh+2012} counts deviate at low flux densities, becoming increasingly lower. At the high flux end, the 7C $151$~MHz source counts \citep{McGilchrist+1990} match our counts well. We have fit a power law over the flux density range $15 - 400$~mJy and obtain a slope of $0.93 \pm 0.04$ which is consistent with, but slightly steeper than, the $0.91$ obtained by  \cite{Intema+2011} across the same flux density range. Likewise, it is consistent with the value of $1.01$ found by \cite{IshwaraChandra+2010}, but is slightly shallower. The source counts derived from the small sample of \cite{GeorgeStevens2008} (not plotted) are fit by a single power law with a slope of $0.72$, but their deviation is probably due to poor statistics. Model source counts have been derived by \cite{Wilman+2008} for the $151$~MHz source population predicted from the extrapolated radio luminosity functions of different radio sources in a $\Lambda$CDM framework. The \cite{Wilman+2008} model catalogue has been corrected with their recommended post-processing, which effectively reduces the source count slightly at low flux densities. The dominant source population at flux densities above $\sim\!200$~mJy is that of FRII radio sources. Only below this flux density does the FRI population begin to dominate. There is a general agreement between our data and this model which has an approximate power-law slope of $0.79$ between $10$ and $400$~mJy. At low flux densities it is likely that the \cite{Wilman+2008} counts slightly overestimate the true counts due to double counting of hybrid AGN-star forming galaxies.
}

\subsection{{Spectral Index Distributions}}
\label{sect:spec_ind}
While deep $1.4$~GHz data exists for the Bo\"{o}tes Field \citep{deVries+2002}, this only covers the central $7$~deg$^2$. This data was used in \cite{Intema+2011} in a $153$\ MHz flux-limited spectral index analysis. However, we choose to compare  our source list to the NVSS $1.4$~GHz catalogue \citep{Condon+1994} which covers our entire survey area at a comparable resolution. We searched for NVSS counterparts within $45\arcsec$ of each GMRT source. Despite the relatively small difference in resolution between the NVSS ($45\arcsec$) and the GMRT ($25\arcsec$) data, a small number of GMRT sources ($9$ pairs) were matched the same NVSS source. Also, due to differences in the grouping of components into sources, we merged $16$ pairs of NVSS sources which matched a single GMRT source. Sources were merged by summing their total flux densities. A spectral index was calculated for each GMRT source based on the combined flux density of merged sources. 

\begin{figure}
 \centering
\resizebox{\hsize}{!}{\includegraphics{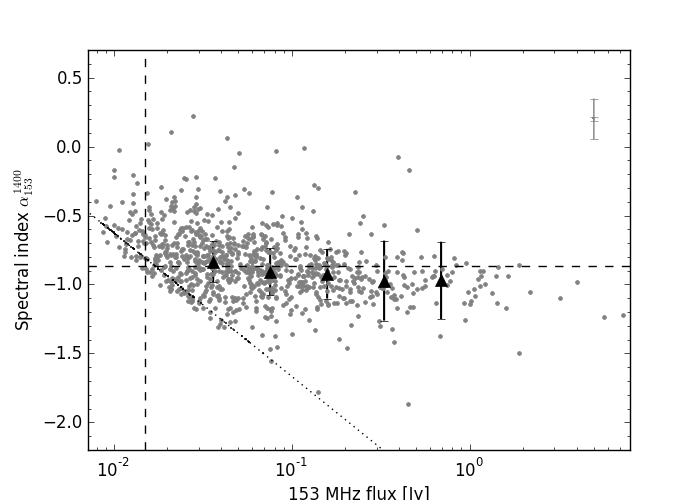} }\\
\caption{Spectral index, $\alpha^{1400}_{153}$, distribution of sources matched between $1.4$~GHz and $153$~MHz (grey points). The  difference in resolution is  $45\arcsec$ (NVSS) and $25\arcsec$ (GMRT) and multiple GMRT matches to a single NVSS source have been merged into one. The vertical line shows $5 \sigma_{avg}$, where $\sigma_{avg}$ is the average $rms$ noise in the GMRT mosaic. The diagonal dotted line indicates the incompleteness limit due to the sensitivity of NVSS and sources with upper limits are plotted as black points along this line. The horizontal dashed line shows the KM mean spectral index of $-0.87 \pm 0.01$ {accounting for upper limits}. The large black triangles show the mean spectral index in $5$ logarithmic bins. Error bars on individual points are not plotted for clarity, but a single bar in the top right indicates the maximum and minumum errors in the dataset.}
\label{fig:specind_nvss}
\end{figure}

We matched  $1134$ NVSS sources to $1127$ GMRT sources and then used this matched subsample to compute the spectral index\footnote{The spectral index is defined as $S_{\nu} \propto \nu^\alpha$} distribution which is shown in Fig.~\ref{fig:specind_nvss}. The flux density limit of $2.5$~mJy at $1.4$~GHz biases the detection of $1.4$~GHz counterparts to fainter $153$~MHz sources to those with flatter spectra. $168$ GMRT sources have no match in NVSS. These are consistent with having steeper spectral indices below the diagonal line in Fig.~\ref{fig:specind_nvss} and we therefore provide an an upper limit to the spectral index given the NVSS flux density limit.  {The mean spectral index is $-0.87 \pm 0.01$, calculated using the Kaplan-Meier estimator \citep[KM; e.g.][]{FeigelsonNelson1985}} to account for the upper limits. This value is comparable to those found by \cite{Intema+2011}, $-0.79$, \cite{IshwaraChandraMarathe2007}, $-0.85$, \cite{Sirothia+2009}, $-0.82$,  and \cite{IshwaraChandra+2010}, $-0.78$. {By considering the KM mean spectral index within $5$ logarithmic flux density bins between $85$~mJy and $1$~Jy (overplotted in Fig.~\ref{fig:specind_nvss} and listed in Table~\ref{tab:specind_nvss_wenss}), we find a gradual steepening of the spectral index with increasing flux density, from $\sim\!-0.84$ at $\sim\!30$~mJy to $\sim\!-0.97$ at $F \gtrsim 600$~mJy.} This trend is still clear if the first flux density bin is ignored (i.e. considering $F \gtrsim 40$~mJy) assuming that this bin remains biased by the upper limits. This is consistent with what is found in the literature \citep[e.g.][]{IshwaraChandra+2010,Tasse+2006,Cohen+2004,deVries+2002}. Since it appears that there is no spectral steepening or flattening due to redshifted curved spectra \citep{Bornancini+2010}, this flattening is likely due to a correlation between source luminosity and spectral index ($P - \alpha$), which is known to exist for FRII radio galaxies \citep[e.g.][]{Blundell+1999}. According to the models of \cite{Wilman+2008}, the observed $153$~MHz source population is dominated by FRII galaxies at these flux density levels ($\gtrsim 20$~mJy).

\begin{table*}
 \begin{center}
 \caption{{Binned median spectral indices between the GMRT at $153$ MHz and NVSS at $1.4$ GHz and WENSS at $327$ MHz).}}
 \label{tab:specind_nvss_wenss}
\begin{tabular}{ccccc}
\hline
\multicolumn{1}{c}{~} & \multicolumn{2}{c}{NVSS $\alpha^{1400}_{153}$}& \multicolumn{2}{c}{WENSS $\alpha^{327}_{153}$}\\
\multicolumn{1}{c}{Bin Centre} & \multicolumn{1}{c}{Counts}& \multicolumn{1}{c}{KM Mean Spectral Index}   & \multicolumn{1}{c}{Counts} & \multicolumn{1}{c}{KM Mean Spectral Index}  \\
\multicolumn{1}{c}{[mJy]} & \multicolumn{1}{c}{(Upper limits)}  &  & \multicolumn{1}{c}{(Upper limits)} &  \\
\hline
36  &  275(10) & $-0.835 \pm 0.015$ &  -  & - \\
76 &  213(7) & $-0.909 \pm 0.017$ &  209(16)  & $-0.940 \pm 0.031$  \\
158 & 143 & $-0.922 \pm 0.018$&  140  & $ -0.752 \pm 0.028 $ \\
331 & 71 &  $-0.972 \pm 0.029$ &  72 &  $-0.792 \pm 0.028$ \\
692  & 28 & $-0.970 \pm 0.028$ &  32  &  $-0.845 \pm 0.033$\\
\hline
 \end{tabular}
 \end{center}
\end{table*}

\begin{figure}
 \centering
\resizebox{\hsize}{!}{\includegraphics{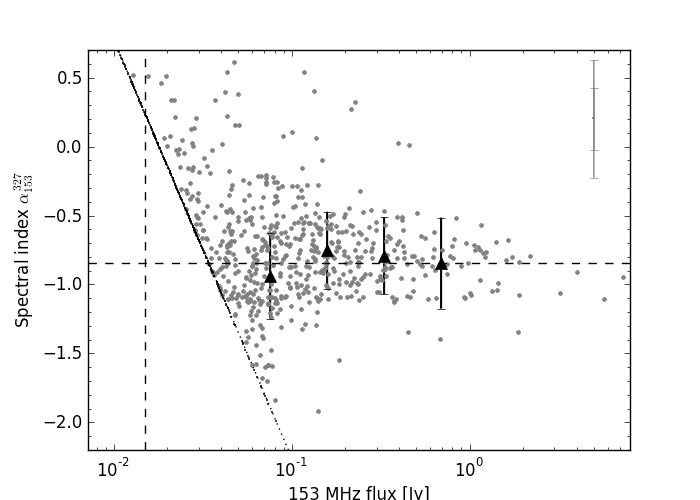} }\\
\caption{Spectral index, $\alpha^{327}_{153}$, distribution between $327$~MHz and $153$~MHz (grey points). The  difference in resolution is  $54\arcsec \times 96\arcsec$ (WENSS) and  $25\arcsec$ (GMRT) and multiple GMRT matches to a single WENSS source have been merged. The vertical line shows $5 \sigma_{avg}$, where $\sigma_{avg}$ is the average $rms$ noise in the GMRT mosaic. The diagonal dotted line indicates the incompleteness limit due to the sensitivity of WENSS. The horizontal dashed line shows the KM mean spectral index of $-0.84 \pm 0.02$ { which takes the upper limits into account}. The large black triangles show the median spectral index in $4$ logarithmic bins. Error bars on individual points are not plotted for clarity, but a single bar in the top right indicates the maximum and minumum errors in the dataset. }
\label{fig:specind_wenss}
\end{figure}

We also compared our source list to WENSS at $327$~MHz \citep{Rengelink+1997}, noting that the errors in this spectral index are much greater due to the smaller difference in frequency. The WENSS beam is $54\arcsec \times 54\arcsec/\sin \delta$, or $54\arcsec \times 96\arcsec$ at the declination of the Bo\"{o}tes field. We thus searched for WENSS counterparts within $96\arcsec$ of each GMRT source. Of the $1289$ GMRT sources we matched $689$ to $675$ WENSS sources. The $14$ pairs of GMRT sources within the beam of a single WENSS source were combined as described in the previous paragraph and spectral indices determined for each based on the combined flux density. A visual check led to the removal of $12$ misidentified or confused sources. The resulting spectral index distribution for $\alpha^{327}_{153}$  is shown in Fig.~\ref{fig:specind_wenss}. Once again there is a bias towards flatter or inverted spectra at low $153$~MHz flux densities due to the WENSS flux density limit of $18$~mJy at $327$~MHz.  We provide upper limits to the spectral indices given the WENSS flux density limit for the $576$ GMRT sources with that have WENSS flux densities below the WENSS detection limit and thus should have spectral indices are steeper than the diagonal line in Fig.~\ref{fig:specind_wenss}. The KM mean spectral index in this case is $-0.84 \pm 0.02$ (taking into account the upper limits) and is {slightly shallower} than that observed between $153$ and $1400$\ MHz. The KM mean spectral indices measured in $4$ flux density bins are also listed in Table~\ref{tab:specind_nvss_wenss}. There is, however, no {clear} trend with flux density observed{, although there is an indication of a slight flattening of the average radio spectrum if the first flux bin is excluded}. This may be due to the fact that  $\alpha^{327}_{153}$ is less robust due to the small frequency difference {and the errors on the individual measurements are higher}.

\begin{figure}
 \centering
\resizebox{\hsize}{!}{\includegraphics{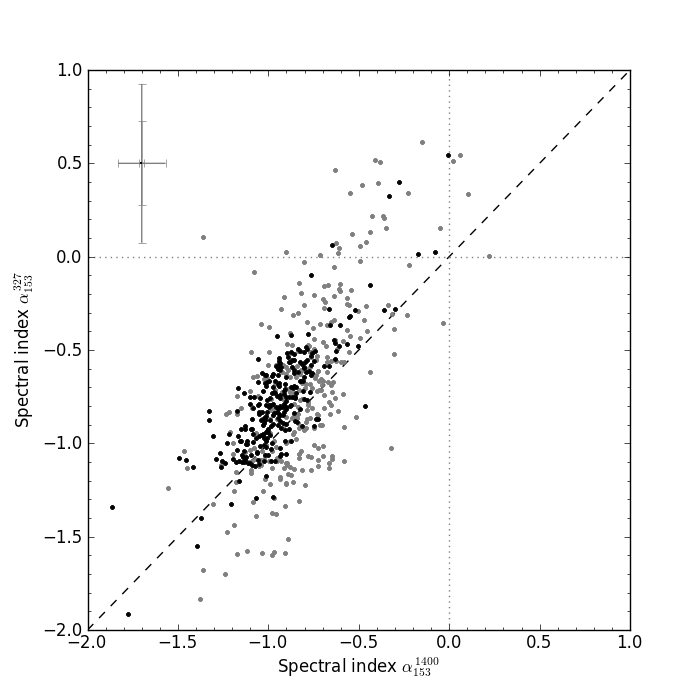} }\\
\caption{Comparison between $\alpha^{1400}_{327}$ and $\alpha^{327}_{153}$. The black dashed line indicates where the spectral index is the same in both regions of the spectrum. Sources with GMRT fluxes above $0.1$~Jy~beam$^{-1}$ are plotted in black and fainter sources are plotted in grey. Error bars on individual points are not plotted for clarity, but a single error bar in the top left indicates the maximum and minumum errors in the dataset.}
\label{fig:speccolcol}
\end{figure}

\begin{figure}
 \centering
\resizebox{\hsize}{!}{\includegraphics{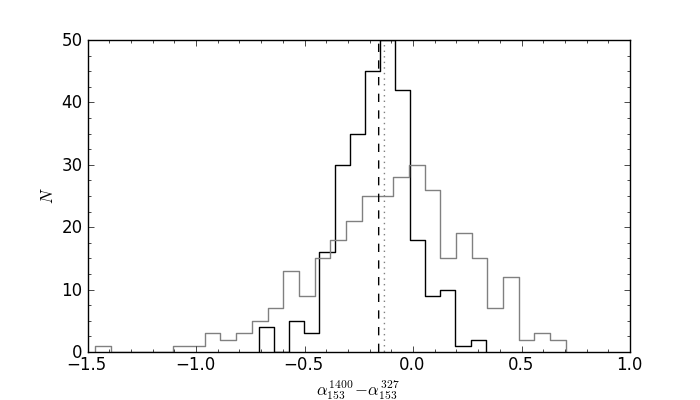} }\\
\caption{Comparison between $\alpha^{1400}_{153}$ and $\alpha^{327}_{153}$: histogram of $\alpha^{1400}_{153}-\alpha^{327}_{153}$. Again, the histogram for bright GMRT sources is plotted in black and for fainter sources in grey. The dashed black line shows the mean value of $-0.25$ and the grey dotted line, the mean value of $-0.2$, for bright and all sources respectively. indicating that the majority of sources have flattened spectra at low frequencies.}
\label{fig:speccolcolhist}
\end{figure}

Around $50$~per~cent of our sources have data at three frequencies ($1400$, $327$, and $153$~MHz), thus we have not attempted to fit or locate peaks in the radio spectra. Instead we show a radio ``colour-colour'' plot, Fig.~\ref{fig:speccolcol}, comparing the spectral indices $\alpha^{1400}_{153}$ and $\alpha^{327}_{153}$. The line illustrates where the two spectral indices are equal. Here we have plotted separately bright $153$~MHz sources, above $0.1$~Jy~beam$^{-1}$, as these sources have smaller errors on their spectral indices and are not affected by incompleteness at the other two frequencies (see Figs.~\ref{fig:specind_nvss} and \ref{fig:specind_wenss}). In general there is a flattening of the average radio spectrum toward lower frequencies, as the majority of points fall above the line. It is likely that this observed turnover in the spectra at low frequencies is due synchrotron self-absorption.  We also plot the distribution of the difference in spectral indices, $\alpha^{1400}_{153}-\alpha^{327}_{153}$, Fig.~\ref{fig:speccolcolhist}, which shows a mean value of $-0.25$  for only bright sources and  and $-0.2$ for all sources.


\begin{figure}
 \centering
\resizebox{\hsize}{!}{\includegraphics{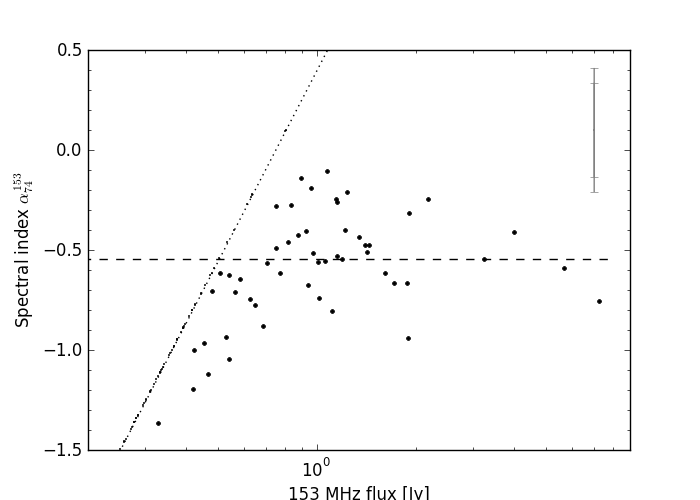}} \\
\caption{Spectral index distribution between $74$ MHz and $153$ MHz. The  difference in resolution is  $80\arcsec$ (VLSS) and the $25\arcsec$ (GMRT) and multiple GMRT matches to a single VLSS source have been merged into one. The vertical line shows $5 \sigma_{avg}$, where $\sigma_{avg}$ is the average $rms$ noise in the GMRT mosaic. The diagonal dotted line indicates the incompleteness limit due to the sensitivity of VLSS. The horizontal dashed line shows the {KM mean} spectral index of $-0.55$. Error bars on individual points are not plotted for clarity, but a single bar in the top right indicates the maximum and minumum errors in the dataset.}
\label{fig:specind_vlss}
\end{figure}

Finally, we have also compared our source list to VLSS at $74$~MHz \citep{Cohen+2007}, again noting that the errors in this spectral index will be much greater due to the smaller difference in frequency. VLSS has a resolution of  $80\arcsec$ so we searched for VLSS sources within this radius of each GMRT source. $58$ GMRT sources were matched to $55$ VLSS sources. The resulting spectral index distribution  is shown in Fig.~\ref{fig:specind_vlss}. In this case these is a bias towards steeper spectra at low $153$~MHz flux densities due to the VLSS flux density limit of $0.5$~Jy at $74$~MHz. The KM mean spectral index in this case is $-0.55$ which was calculated for sources with GMRT fluxes above $0.5$~Jy.


%
%
%

\section{Conclusion}
\label{sect:concl}
We have presented the results from a  $\sim\!30$ square degree,  high resolution ($25\arcsec$) radio survey at $153$\ MHz centred on the NOAO Bo\"{o}tes field. We have employed the SPAM ionospheric calibration scheme to achieve an $rms$ noise in the $7$ pointing mosaicked image of $\sim\!2 - 4$\ mJy\ beam$^{-1}$. The source catalogue contains $1289$ sources between $4.1$\ mJy and $7.3$\ Jy detected at $5$ times the local noise. We estimate the catalogue to be $92$~per~cent reliable and $95$~per~cent complete to an integrated flux density of $14$~mJy. The catalogue has been corrected for systematic errors on both the astrometry and flux density scales.

We have analysed the source population by investigating the source counts and by identifying counterparts within the $1.4$~GHz NVSS and $327$~MHz WENSS surveys and have computed the spectral index distributions of these sources. Understanding the low frequency, low flux source population is of particular importance to Epoch of Reionization projects \citep[e.g.][and references therein]{Ghosh+2012} where good models of the foregrounds are needed.

In the near future, this data will be combined with the existing multi-wavelength data covering the NOAO Bo\"{o}tes field and we will study the  properties of radio galaxies as a function of various multi-wavelength parameters across a range of cosmic time. Further investigation of the spectral indices will be done and can be used to identify USS sources as well as high redshift gigahertz peaked spectrum (GPS) sources.

\begin{acknowledgements}
The authors thank the anonymous referee for useful comments which have improved this manuscript. We also acknowledge the staff of the GMRT that made these observations possible. GMRT is run by the National Centre for Radio Astrophysics of the Tata Institute of Fundamental Research. This publication made use of data from the Very Large Array, operated by the National Radio astronomy Observatory. The National Radio Astronomy Observatory is a facility of the National Science Foundation operated under cooperative agreement by Associated Universities, Inc.
\end{acknowledgements}

\bibliographystyle{aa}
\bibliography{bibfile}

\appendix
\section{Selected Radio Images}
\label{app:images}
Figures \ref{fig:postage} shows the $25$ brightest sources in the catalogue, excluding the second brightest source which is described below (see also Fig.\ \ref{fig:special_bright}). 

\subsection{Note on source J144102+3530}
\begin{figure}
 \centering
\resizebox{0.8\hsize}{!}{\includegraphics{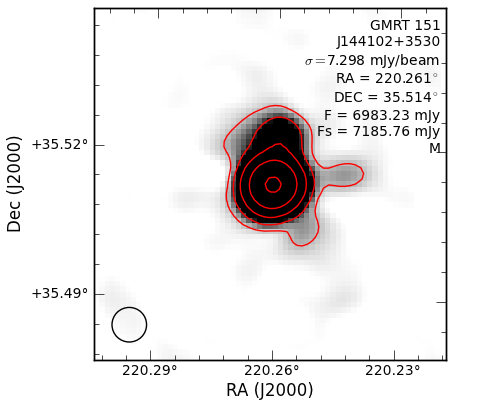} }\\
\resizebox{0.8\hsize}{!}{\includegraphics{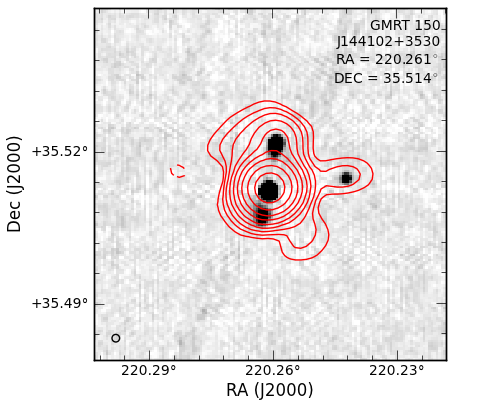} }\\
\caption{The second brightest source in the $153$\ MHz catalogue: (top) GMRT image and (bottom) FIRST image with GMRT contours. In both images the GMRT countours are plotted in red at intervals of $3\sigma \times [-\sqrt{3}, \sqrt{3}, \sqrt{10}, \sqrt{30}, \sqrt{100}, \ldots]$ and the greyscale goes from $1\sigma$ to $30\sigma$.}
\label{fig:special_bright}
\end{figure}

Figure \ref{fig:special_bright} shows GMRT postage stamp of the second brightest source in the catalogue. Also shown is the  FIRST image \citep{Becker+1995} of this source which shows that most of the structure seen in the GMRT image is in fact real. Only the extension to the North-West in the GMRT image has no clear match in the FIRST image and may be due to deconvolution errors.

\begin{figure*}
\centering
\includegraphics[width=0.32\textwidth]{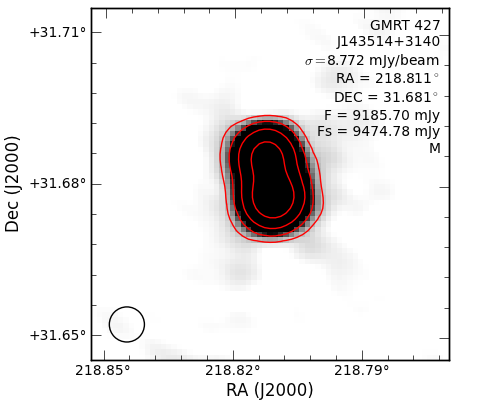}
\includegraphics[width=0.32\textwidth]{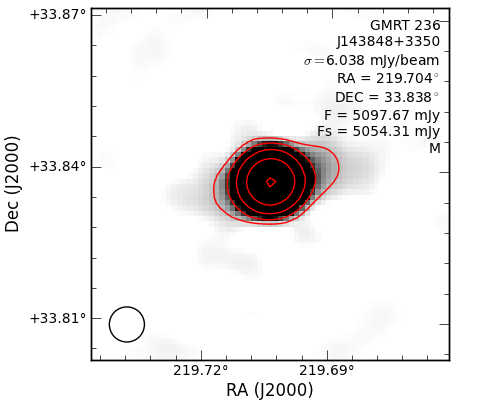}
\includegraphics[width=0.32\textwidth]{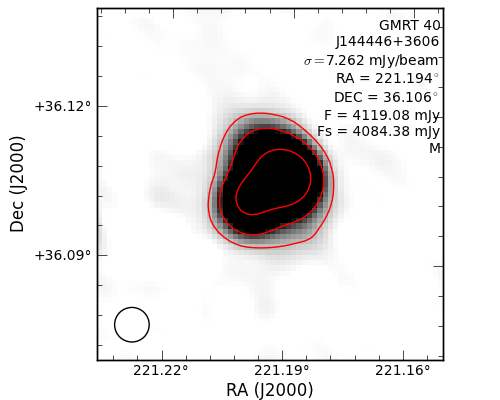}\\
\includegraphics[width=0.32\textwidth]{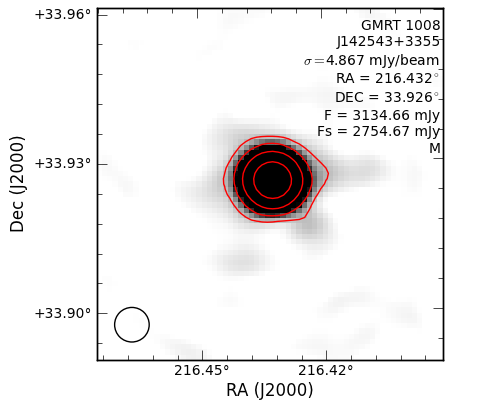}
\includegraphics[width=0.32\textwidth]{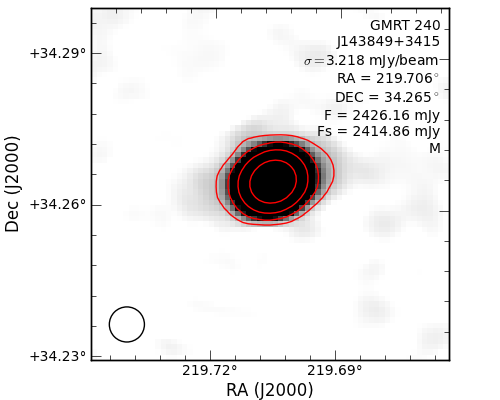}
\includegraphics[width=0.32\textwidth]{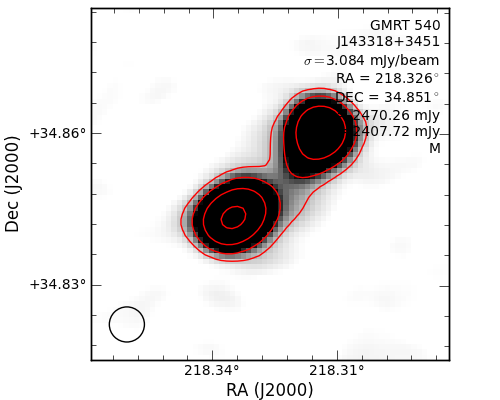}\\
\includegraphics[width=0.32\textwidth]{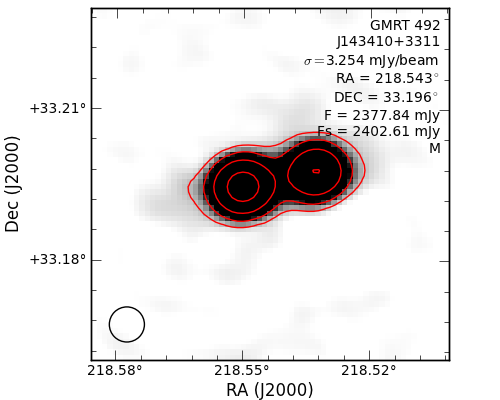}
\includegraphics[width=0.32\textwidth]{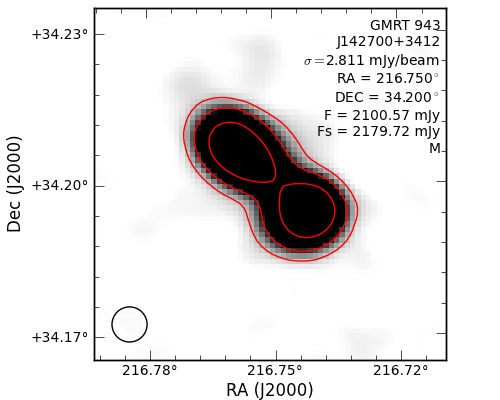}
\includegraphics[width=0.32\textwidth]{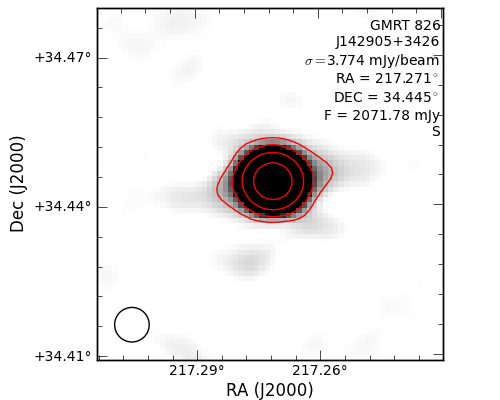}\\
\includegraphics[width=0.32\textwidth]{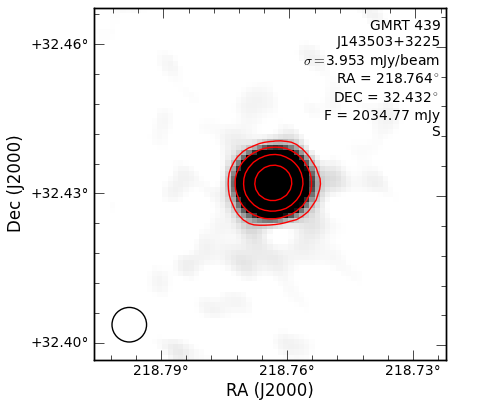}
\includegraphics[width=0.32\textwidth]{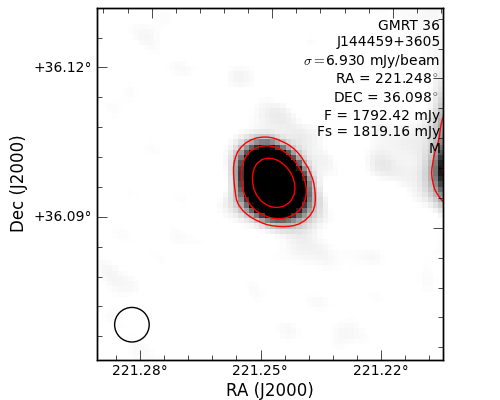}
\includegraphics[width=0.32\textwidth]{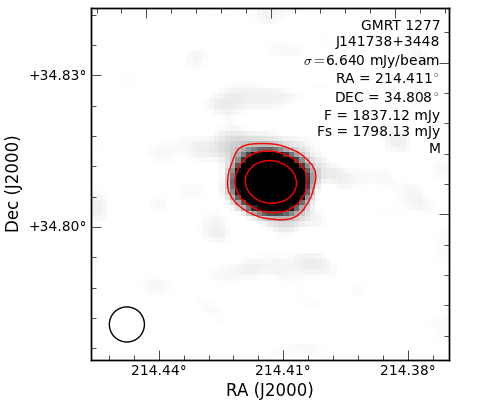}\\

\caption{The $25$ brightest $153$~MHz radio sources (exluding J144102+3530). Countours are plotted in red at intervals of $3\sigma \times (-\sqrt{3}, \sqrt{3}, \sqrt{10}, \sqrt{30}, \sqrt{100}, \ldots)$ and the greyscale goes from $1\sigma$ to $30\sigma$.  The text in each image lists the local $rms$ noise, the source coordinates and total flux, density and the source type (`S' or `M'). The beamsize is shown in the bottom left corner.}
\label{fig:postage}
\end{figure*}
\begin{figure*}
\ContinuedFloat
\includegraphics[width=0.32\textwidth]{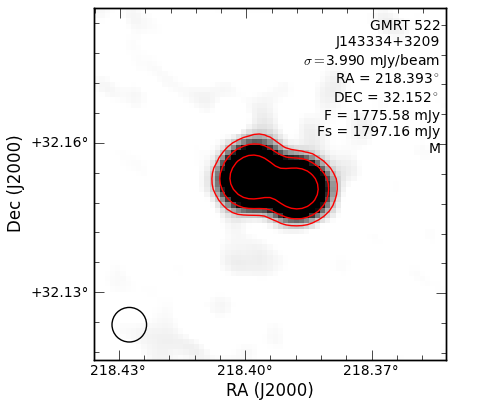}
\includegraphics[width=0.32\textwidth]{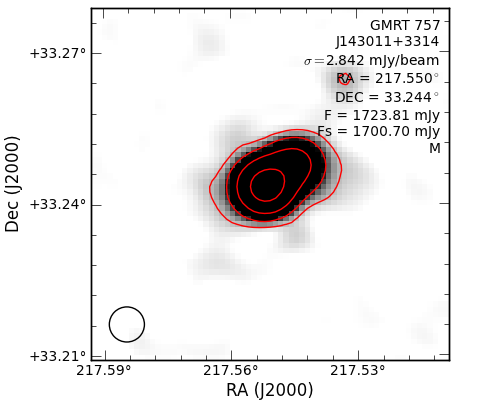}
\includegraphics[width=0.32\textwidth]{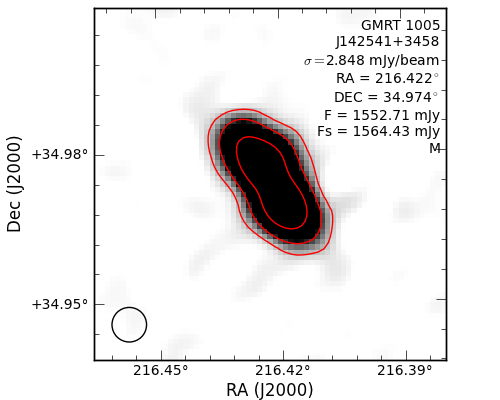}\\
\includegraphics[width=0.32\textwidth]{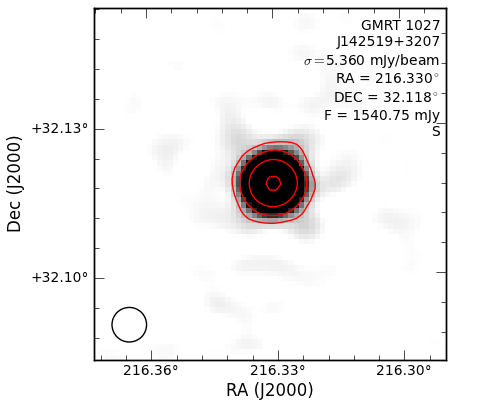}
\includegraphics[width=0.32\textwidth]{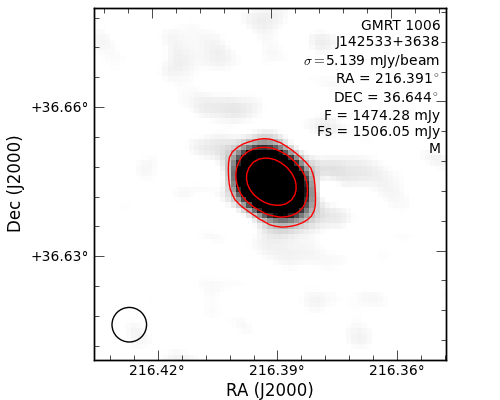}
\includegraphics[width=0.32\textwidth]{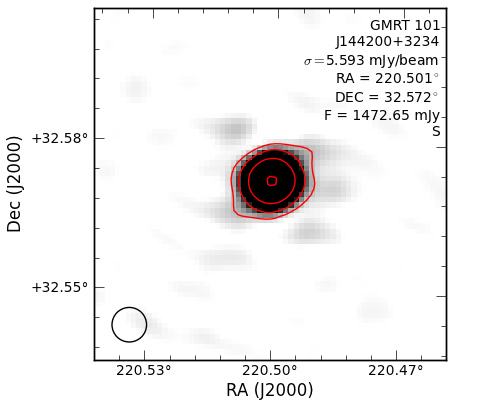}\\
\includegraphics[width=0.32\textwidth]{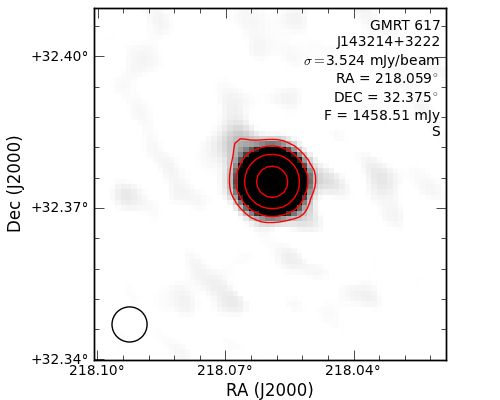}
\includegraphics[width=0.32\textwidth]{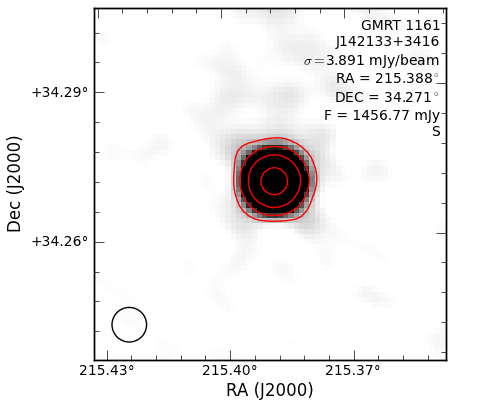}
\includegraphics[width=0.32\textwidth]{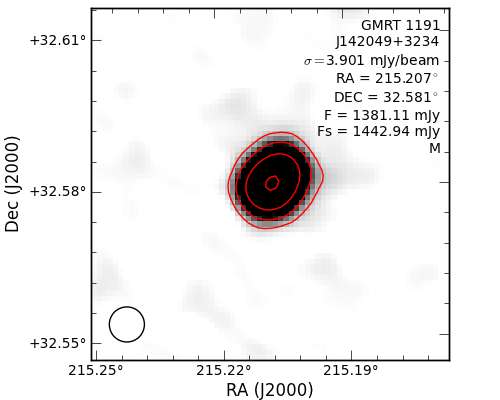}\\
\includegraphics[width=0.32\textwidth]{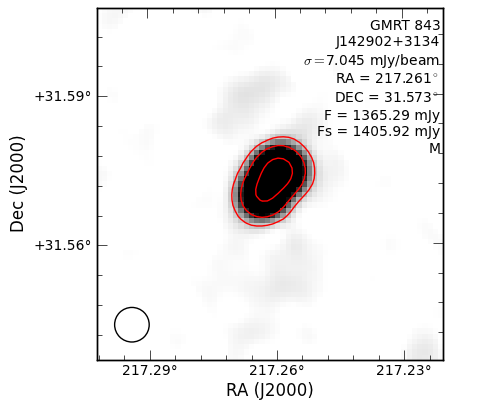}
\includegraphics[width=0.32\textwidth]{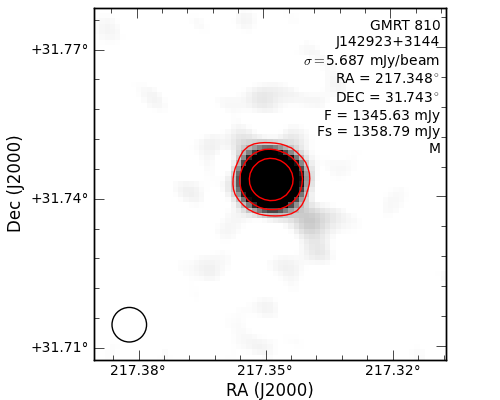}
\includegraphics[width=0.32\textwidth]{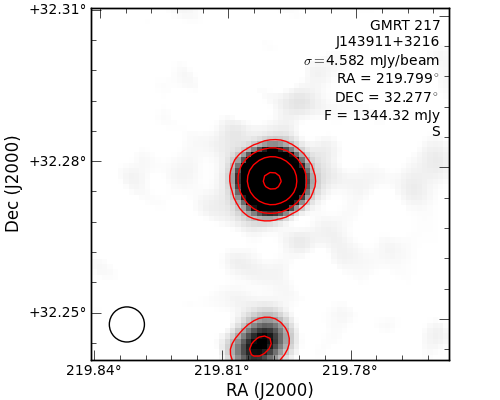}\\
 \caption{Continued.}
\end{figure*}

\end{document}